\documentclass[12pt]{article}
\usepackage{epsfig,amssymb,amsmath,psfrag}

\catcode`\@=11

\def \bl  {\begin{align*}}
\def \el  {\end{align*}}

\def \be  {\begin{equation}}
\def \ee  {\end{equation}}
\def \ba  {\begin{eqnarray}}
\def \ea  {\end{eqnarray}}
\def \baa {\begin{eqnarray*}}
\def \eaa {\end{eqnarray*}}
\def \bb  {\begin {thebibliography} }
\def \eb  {\end{thebibliography}}
\def \lab #1 {\label{#1}}

\newcommand\re[1]{(\ref{#1})}
\def \qqquad {\qquad\quad}

\def \matrix #1 {\left(\begin{array}{cc} #1 \end{array}\right)}

\def \tr {\mathop{\rm tr}\nolimits}

\def \e  {\mathop{\rm e}\nolimits}
\newcommand\lr[1]{{\left({#1}\right)}}
\newcommand \widebar [1] {\overline{#1}}

\newcommand \vev [1] {\langle{#1}\rangle}

\newcommand \ket [1] {|{#1}\rangle}
\newcommand \bra [1] {\langle {#1}|}

\newcommand{\ft}[2]{{\textstyle\frac{#1}{#2}}}

\newcommand{\cN}{{\cal N}}
\newcommand{\p}[1]{(\ref{#1})}

\def\XXint#1#2#3{{\setbox0=\hbox{$#1{#2#3}{\int}$}
     \vcenter{\hbox{$#2#3$}}\kern-.5\wd0}}

\def\l<{\langle}\def\r>{\rangle}
\def\bMHV{\overline{\rm MHV}}

\def\numberbysection{\@addtoreset{equation}{section}
                     \def\theequation{\thesection.\arabic{equation}}}
\numberbysection

\begin{document}

\thispagestyle{empty}
\null\vskip-12pt \hfill  LAPTH--1264/08 \\
\null\vskip-12pt \hfill LPT--Orsay--08--72
\vskip2.2truecm
\begin{center}
\vskip 0.2truecm {\Large\bf
{\Large Generalized unitarity for $\mathcal{N}=4$ super-amplitudes}
}\\
\vskip 1truecm
{\bf J.M. Drummond$^{*}$, J. Henn$^{*}$, G.P. Korchemsky$^{**}$ and E. Sokatchev$^{*}$ \\
}

\vskip 0.4truecm
$^{*}$ {\it
LAPTH\footnote{Laboratoire d'Annecy-le-Vieux de Physique Th\'{e}orique, UMR 5108}, Universit\'{e} de Savoie, CNRS\\
B.P. 110,  F-74941 Annecy-le-Vieux Cedex, France\\
\vskip .2truecm $^{**}$ {\it
Laboratoire de Physique Th\'eorique%
\footnote{Unit\'e Mixte de Recherche du CNRS (UMR 8627)},
Universit\'e de Paris XI, \\
F-91405 Orsay Cedex, France
                       }
  } \\
\end{center}

\vskip 1truecm 
\centerline{\bf Abstract} 

\medskip

We develop a manifestly supersymmetric version of the generalized unitarity cut method for
calculating scattering amplitudes in $\mathcal{N}=4$ SYM theory. We illustrate the power of this
method by computing the one-loop $n$-point NMHV super-amplitudes. The result confirms two
conjectures which we made in arXiv:0807.1095 [hep-th]. Firstly, we derive the compact, manifestly
dual superconformally covariant form of the  NMHV tree amplitudes for arbitrary number and types of
external particles. Secondly, we show that the ratio of the one-loop NMHV to the MHV amplitude is
dual conformal invariant.

\newpage
\setcounter{page}{1}\setcounter{footnote}{0}

\tableofcontents
\newpage


\section{Introduction}

In this paper, we continue the study of a new symmetry of scattering amplitudes in $\mathcal{N}=4$
supersymmetric Yang-Mills theory (SYM),  dual superconformal symmetry \cite{dhks5}. This symmetry
goes beyond all known symmetries of $\mathcal{N}=4$ theory and it was conjectured in \cite{dhks5} to
hold both at weak and at strong coupling. It relates various particle amplitudes with different
helicity configurations (maximally helicity violating (MHV), next-to-MHV and so on) and imposes
non-trivial constraints on the loop corrections. Dual superconformal symmetry is related through the
AdS/CFT correspondence to the invariance of the sigma-model on the ${\rm AdS}_5\times {\rm S}^5$
background under T-duality transformations applied to both bosonic \cite{Kallosh:1998ji} and
fermionic \cite{Berkovits:2008ic,Beisert:2008iq} variables.

Calculating scattering amplitudes in $\mathcal{N}=4$ SYM is a complicated task, due to the
large variety of scattered on-shell states and their helicity configurations, and due to the
number of contributing diagrams in the perturbative expansion. To uncover the dual superconformal
symmetry of the scattering amplitudes we need an efficient method for computing them in
$\cN=4$ SYM that takes full advantage of supersymmetry of the underlying gauge theory.
In this paper, we develop such a method and apply it to derive the dual superconformal covariant
representation of the MHV and NMHV super-amplitudes conjectured in \cite{dhks5} .

The method represents a manifestly $\cN=4$ supersymmetric version of the generalised unitarity cut method.
Unitarity of the S-matrix is a property which has been exploited widely in quantum field theory. It
is a particularly strong constraint on the perturbative structure of scattering amplitudes in
supersymmetric gauge theories. Indeed in such theories the amplitudes can be argued to be
cut-constructible, that is they can be entirely determined by a knowledge of the structure of their
branch cuts, as shown in \cite{bddk94,bddk94-2}. In these papers it is shown that in
one-loop calculations in $\cN=4$ SYM  all integrals can be reduced
\cite{integral-reduction} to one-loop box integrals. Thus, a one-loop amplitude in $\cN=4$ SYM is
completely determined by the rational prefactors of the box integrals. The latter can be determined
by considering cuts that give the discontinuities of the integral functions corresponding to
particular kinematical invariants. At one loop, this amounts to cutting two propagators.

A generalisation of this method consisted in taking cuts passing through three or more propagators
\cite{Bern:1997sc}. At one loop, the most restrictive of these cuts is the four-particle cut which
fixes the loop integration completely. This is very convenient since it allows to compute one box
coefficient at a time, and because no phase space integration is required. There is a technical
difficulty, however; if one works in Minkowski signature such four-particle cuts only give
information on the coefficients of four-mass box integrals, because for the others there will always
be a three-point on-shell vertex which vanishes. A way around this obstruction to using the
four-particle cut was found in \cite{bcf1}; one can either replace the signature by $(++--)$, or
complexify the momenta, which allows a non-zero on-shell three-point vertex. The generalised cut
technique could then be applied to the computation of various one-loop amplitudes in $\cN=4$ SYM,
namely the $n$-gluon MHV and NMHV amplitudes for arbitrary $n$ \cite{bddk94,Bern:2004bt}. Amplitudes
with other external particles were also computed, see e.g. \cite{Risager:2005ke}.

In gauge theories in general it is often necessary to consider different distributions of the
external particle and their helicities and also to sum over all possible intermediate states.
In supersymmetric gauge theories both of these considerations can be neatly combined by using a
superspace representation for the relevant amplitudes \cite{Nair:1988bq}. The external
particle/helicity choices are then kept general as they are encoded in the superspace structure and
the sum over intermediate states is simply replaced by a Grassmann integration over the superspace
coordinates of the various sub-amplitudes separated by the cuts. Amplitudes in superspace were
already considered in \cite{Georgiou:2004by,bst}, although in a different context.

In this paper we will give some non-trivial examples of using superspace and unitarity to obtain a
compact form for amplitudes in $\cN=4$ SYM. The methods described here should easily generalise to
$\cN=8$ supergravity. We first illustrate the method on the known expressions for one-loop $n$-point
MHV amplitudes \cite{bddk94} using generalised cuts. Our main new result is the computation of the
$n$-point NMHV super-amplitudes at one-loop (note that partial results for $6$-point NMHV amplitudes
using similar ideas were already obtained in \cite{Huang}). The result confirms the proposal we made
in \cite{dhks5}, and also agrees with previously known results for scattering amplitudes involving
gauge fields, gaugino and scalars \cite{Bern:2004bt,Risager:2005ke}. By infrared consistency of the
one-loop amplitudes we derive a compact expression for the $n$-point NMHV tree-level
super-amplitudes (also conjectured in \cite{dhks5}).

Writing the amplitudes in superspace also allows us to discover their dual superconformal
properties. Dual superconformal symmetry was introduced in \cite{dhks5} as a generalisation of dual
conformal symmetry \cite{dhss,am1,dks,dhks1,dhks2}. The latter is an important aspect of the
remarkable duality between scattering amplitudes and Wilson loops in $\mathcal{N}=4$ SYM theory
\cite{am1,dks,bht}. In this paper we obtain the manifestly dual superconformal form of the three-,
two- and one-mass box super-coefficients.~\footnote{The three-mass box case were already presented
in \cite{dhks5} based on comparison with the $n$-gluon NMHV amplitudes from \cite{Bern:2004bt}.} The
NMHV tree-level super-amplitude that we find is made of three-mass box coefficients and hence is
also manifestly dual superconformally covariant. In a recent paper \cite{Brandhuber:2008pf} a
recursive argument was used to check the conjecture made in \cite{dhks5} that all tree-level
$\mathcal{N}=4$ SYM super-amplitudes have this property.

The paper is organised as follows. In section \ref{sect-superspace}, we introduce the necessary
notations and concepts to describe scattering amplitudes in $\cN=4$ SYM in an on-shell superspace.
In section \ref{sect-unitarity-MHV} we show how unitarity cuts can be evaluated in superspace on the
known example of $n$-point MHV super-amplitudes. In section \ref{section-unitarity-NMHV} we present new
results for $n$-point NMHV super-amplitudes. We use these results to show the dual superconformal
properties of the latter in section \ref{sect-superdual}.

\section{Scattering amplitudes in $\mathcal{N}=4$ SYM}
\label{sect-superspace}

In this section we briefly review the structure of scattering amplitudes in $\mathcal{N}=4$ super
Yang-Mills theory. The on-shell states in this theory are two gluons $G^\pm$ with helicity $\pm 1$,
eight gluinos $\Gamma_A, \bar{\Gamma}^A$ with helicity $\pm\tfrac 12$ and six real scalars described
by a complex wave function $S_{AB}$ satisfying the reality condition $S_{AB} = \tfrac{1}{2}
\epsilon_{ABCD}\bar{S}^{CD}$. Here the upper (lower) indices $A,B,C,D=1,\ldots,4$ correspond to the
(anti-) fundamental representation of the R symmetry group $SU(4)$ of the $\mathcal{N}=4$ theory.

We can label the on-shell states by $(p_i,h_i,a_i)$ with $h_i$ being the helicity, $a_i$ the color
index in the adjoint representation of the gauge group $SU(N)$ and $p_i^\mu$ the light-like momentum
of the $i$-th particle ($p_i^2=0$). The generic $n-$particle scattering amplitude in the
planar $\mathcal{N}=4$ theory then has the following form
\begin{align}\label{A-planar}
\mathcal{A}_n(\{p_i,h_i,a_i\}) =(2\pi)^4\delta^{(4)}\big(\sum_{i=1}^n p_i\big)\!\! \sum_{\sigma\in
S_n/Z_n} 2^{n/2}g^{n-2} \tr [t^{a_{\sigma(1)}}\ldots t^{a_{\sigma(n)}} ]
A_n\left(\sigma(1^{h_1},\ldots, n^{h_n})\right) \,,
\end{align}
where the sum runs over all possible non-cyclic permutations $\sigma$ of the set $\{1,\ldots,n\}$
and the color trace involves the $SU(N)$ generators $t^a$ in the fundamental representation
normalized as $\tr(t^a t^b) =\frac12\delta^{ab}$. All particles are treated as incoming, so that
momentum conservation takes the form $\sum_{i=1}^n p_i=0$. A convenient way to describe the
helicities of the various particles is the spinor formalism, in which one solves the on-shell
conditions $p_i^2=0$ by expressing each the light-like momentum in terms of a pair of commuting
spinors,
\be\label{lambda}
(p_i)^{\dot\alpha\alpha} = \tilde\lambda_i^{\dot\alpha}\lambda_i^\alpha\,,
\ee
where $p_i^\mu =\tfrac12 p_i^{\dot\alpha\alpha}\sigma^\mu_{\alpha\dot\alpha}$ and
$\sigma^\mu=({\mathbb I},\vec \sigma)$ is given by the Pauli matrices. In a physically realistic
situation, where the momenta are real vectors in four-dimensional Minkowski space with signature
$(+---)$, the spinor $\lambda$ belongs to the fundamental (chiral) representation of the Lorentz
group $SL(2,{\mathbb C})$,  while its complex conjugate $\tilde\lambda = \bar\lambda$ belongs to the
anti-fundamental (anti-chiral) representation. However, one of the key points in the generalized
unitarity approach that we apply in the present paper,  is the use of three-particle amplitudes,
which do not exist for real Minkowski momenta. One way to introduce them is to complexify the
momenta, and hence the Lorentz group. Therefore we shall often treat $\tilde\lambda$ as yet another
complex two-component spinor, unrelated to $\lambda$ by complex conjugation.

The color-ordered partial amplitudes $A_n\left(\sigma(1^{h_1},\ldots, n^{h_n})\right)$ only depend
on the momenta and helicities of the particles and admit a perturbative expansion in powers of the
`t Hooft coupling $a=g^2N/(8\pi^2)$. The color-ordered amplitudes can be classified according to
their total helicity $h_{\rm tot} = h_1+\ldots+h_n$, whose possible values are $h_{\rm tot} =
-n,-n+2,\ldots,n-2,n$. As a corollary of supersymmetry, the amplitudes with total helicity $h_{\rm
tot} = \pm n,\, \pm (n-2)$ vanish at all perturbative orders. The amplitudes with $h_{\rm
tot}=n-4,n-6,\ldots$ are usually referred to as maximally helicity violating (MHV),  next-to-MHV
(NMHV), $\ldots$ amplitudes. Similarly, the amplitudes with $h_{\rm tot}=-(n-4),-(n-6),\ldots$ are
known as $\rm \overline{MHV}$, $\rm \overline{NMHV}$, $\ldots$ amplitudes.

\subsection{Super-amplitudes}

The large number of species of incoming particles in the $\cN=4$ SYM theory leads to a proliferation
of possible scattering amplitudes. Supersymmetry provides us with a very useful bookkeeping tool for
their description. A unique feature of the $\mathcal{N}=4$ SYM theory is that all on-shell states
can be assembled into a single super-wavefunction $\Phi(p,\eta)$ by introducing Grassmann variables
$\eta^A$ (with $A=1,\ldots,4$) belonging to the fundamental representation of R symmetry group
$SU(4)$,
\begin{eqnarray} \label{fitinto}
  \Phi(p,\eta) &=& G^{+}(p) + \eta^A \Gamma_A(p) + \frac{1}{2}\eta^A \eta^B S_{AB}(p)
  + \frac{1}{3!}\eta^A\eta^B\eta^C \epsilon_{ABCD} \bar\Gamma^{D}(p) \nonumber \\
  &&\  + \frac{1}{4!}\eta^A\eta^B\eta^C \eta^D \epsilon_{ABCD} G^{-}(p)\,.
\end{eqnarray}
Here the first term on the right-hand side is the helicity $+1$ state. In each subsequent term the
helicity of the states decreases by a step of $(-1/2$), so that the last term is the helicity $(-1)$
state. It is logical to assign helicity $(-1/2)$ to the variables $\eta^A$, so that each term on the
right-hand side of Eq.~\re{fitinto} has the same total helicity $(+1)$.

The super-wavefunction $\Phi(p,\eta)$ serves as a generating function for the various species of
scattered particles. Thus, for a given number of external particles $n$, all possible scattering
amplitudes can be obtained as components of a \textsl{single super-amplitude},
\be\label{A=Phi}
{\cal A}_n\big(\lambda,\tilde{\lambda},\eta\big) = \mathcal{A}\left( \Phi_1 \ldots \Phi_n \right)\,,
\ee
where $\Phi_i$ is a shorthand for $\Phi(p_i,\eta_i)$ and the spinors $\lambda_i,\tilde{\lambda}_i$
are defined in \re{lambda}. Expanding the super-amplitude ${\cal
A}_n\big(\lambda,\tilde{\lambda},\eta\big)$ in  the $\eta$'s, we can read off the various scattering
amplitudes as the coefficients of the relevant powers of $\eta$'s. For instance,
\begin{align}
{\cal A}_n\big(\lambda,\tilde{\lambda},\eta\big) & =  \left(\eta_1\right)^4 \left(\eta_2\right)^4
\,\mathcal{A}_n\left(G^-G^-G^+\ldots G^+ \right)
\\ \nonumber
& + \frac{1}{3!} \left(\eta_1\right)^4 \eta_2^A \eta_2^B \eta_2^C \eta_3^E \epsilon_{ABCD}\,
\mathcal{A}_n\left( G^-\,
\bar\Gamma^D_2 \Gamma_{3E} G^+\ldots G^+ \right) + \ldots\,,
\end{align}
with $(\eta)^4 = \tfrac1{4!}\eta^A\eta^B\eta^C\eta^D \epsilon_{ABCD}$.

The $\mathcal{N}=4$ super-multiplet described by \re{fitinto} is PCT self-conjugate and, hence, the
complex conjugate super-wave function $\bar \Phi \equiv \lr{\Phi(p,\eta)}^\ast$ is just a different
representation of the same multiplet,
\begin{eqnarray} \label{antiholo}
  \bar\Phi(p,\bar\eta) &=& G^-(p) + \bar\eta_A \bar\Gamma^A(p) + \frac{1}{2}\bar\eta_A \bar\eta_B \bar S^{AB}(p)
  + \frac{1}{3!}\bar\eta_A \bar\eta_B \bar\eta_C  \epsilon^{ABCD} \Gamma_D(p) \nonumber \\
  &&\  + \frac{1}{4!}\bar\eta_A \bar\eta_B \bar\eta_C \bar\eta_D \epsilon^{ABCD} G^+(p)\,,
\end{eqnarray}
where $\lr{\lambda_i^{\alpha}}^* = \tilde\lambda_i^{\dot\alpha}$, $\lr{G^+}^* = G^{-}$,
$\lr{\Gamma_A}^* = -
\bar\Gamma^A$, and $\bar \eta_A = \lr{\eta^A}^*$ belongs to the anti-fundamental representation of
$SU(4)$. By analogy with the two-component Lorentz spinors, we can call the description \p{fitinto}
chiral (or holomorphic, since only the variables $\eta^A$  appear), and \p{antiholo} anti-chiral (or
antiholomorphic).

Note if we complexify the particle momenta, we cannot treat \p{antiholo} as the complex conjugate of
\p{fitinto} anymore. Nevertheless, the two super-wavefunctions $\Phi(p,\eta)$ and
$\bar{\Phi}(p,\bar\eta)$ are still related to each other through the Grassmann Fourier transform
\be\label{Grassmann-Fourier}
\bar{\Phi}(p,\bar\eta) = \int d^4\eta \,\e^{\eta^A \bar\eta_A} \Phi(p,\eta)\,,
\ee
where $d^4\eta = \prod_{A=1}^4 d\eta^A$ and the Grassmann integration uses the rules $\int d\eta^A
=0$ and $\int d\eta^A\ \eta^A = 1$ (no summation over $A$). We can say that even for complexified
momenta the two alternative descriptions of the $\cN=4$ SYM multiplet are Fourier (if not complex)
conjugate to each other.

By analogy with \re{A=Phi}, we define the conjugate super-amplitude as
\be\label{A=Phi'}
\bar {\cal A}_n\big(\lambda,\tilde{\lambda},\eta\big) = \mathcal{A}\left( \bar\Phi_1 \ldots
 \bar\Phi_n \right) = {\cal A}_n\big(\lambda,\tilde{\lambda},\eta\big)\big|_{\eta_i\to \bar\eta_i,\,
 \lambda_i\to \tilde\lambda_i,\, \tilde\lambda_i\to \lambda_i} \,,
\ee
where in the last relation we followed the rule that the conjugated super-wavefunction is obtained
by substituting the spinors, $\lambda \leftrightarrows \tilde{\lambda}$, and the Grassmann
variables, $\eta \leftrightarrows \bar\eta$. In addition, the transform \re{Grassmann-Fourier} leads
to the following relation between the two super-amplitudes
\begin{align} \label{dual}
{\cal A}_n\big(\lambda,\tilde{\lambda},\eta\big) = \int \prod_{i=1}^n d^4\bar\eta_i
\,\e^{-\eta_i\cdot\bar\eta_{i}}\,\bar {\cal A}_n\big(\lambda,\tilde{\lambda},\eta\big) =  \int
\prod_{i=1}^n d^4\bar\eta_i \,\e^{-\eta_i\cdot\bar\eta_{i}}\,{\cal
A}_n\big(\tilde{\lambda},\lambda,\eta\big)\,,
\end{align}
where the super-amplitude in the second relation is obtained from ${\cal
A}_n\big(\lambda,\tilde{\lambda},\eta\big)$ through the substitution \re{A=Phi}.~\footnote{Except
for the case $n=3$, see Section \ref{trlen3}.}

In this paper we make our choice in favor of the holomorphic description, i.e. we always define the
$n$-particle amplitudes with $\Phi$ everywhere. Equivalently, we could have chosen to represent
some or all of the $n$ particles by $\bar{\Phi}$, since it describes the same supermultiplet.

\subsection{Supersymmetry invariance}\label{syinva}

Let us now discuss the consequences of supersymmetry for the super-amplitudes. One of the advantages
of using the (chiral) super-wavefunctions $\Phi(p,\eta)$ is that the supersymmetry transformations
of the various on-shell states entering \re{fitinto} can be presented in the compact form,
\be\label{Phi-SUSY}
\delta \Phi(p,\eta) = \lr{ \epsilon^\alpha_A \, q_\alpha^A + \bar\epsilon^{A\,\dot\alpha} \, \bar
q_{A\,\dot\alpha }}  \Phi(p,\eta)\,,
\ee
with generators $q_{\alpha}^A = \lambda_{\alpha}\eta^A$ and $\bar{q}_{A\,\dot\alpha} =
\tilde\lambda_{\dot\alpha}\partial_{\eta^A}$. For a super-amplitude ${\cal A}_n\big(
\lambda,\tilde{\lambda},\eta\big)$ depending on $n$ super-wavefunctions $\Phi(p_i,\eta_i)$, the
generators of the $\cN=4$ Poincar\'e supersymmetry algebra are given by the sums of the
single-particle generators
\be\label{q}
q_{\alpha}^A = \sum_{i=1}^n \lambda_{i \alpha} \eta_i^A, \qquad
\bar{q}_{A\,\dot\alpha } = \sum_{i=1}^n \tilde\lambda_{i \dot\alpha}
\frac{\partial}{\partial \eta_i^A}\,, \qquad \{q_{\alpha}^A, \bar{q}_{\dot\alpha B} \} = \delta_B^A
\, p_{\alpha \dot\alpha}\,,
\ee
where $p_{\alpha \dot\alpha} = \sum_{i=1}^n p_{i\, \alpha \dot\alpha} = \sum_{i=1}^n \lambda_{i
\alpha} \tilde\lambda_{i \dot\alpha}$ is the total momentum. The invariance of the super-amplitude
${\cal A}_n(\lambda,\tilde{\lambda},\eta)$ under the supersymmetry transformations \re{Phi-SUSY}
means that it is annihilated by the corresponding generators,
\be\label{SUSY}
 q_{\alpha}^A\,{\cal A}_n  =
\bar{q}_{A\,\dot\alpha }  \,{\cal A}_n  = p_{\alpha \dot\alpha}\,{\cal A}_n  = 0\,.
\ee
These relations imply that in the $\mathcal{N}=4$ SYM theory the super-amplitude takes the following
form \footnote{ This formula is true for $n\ge4$. However, for $n=3$ one can construct e.g.
amplitudes $A_{3}(1^-,2^+,3^+)\neq0$ provided the on-shell momenta are complex \cite{Witten:2003nn}.
In this exceptional case the super-amplitude takes a different form, see Section \ref{trlen3}. },
\be\label{superamplitude}
    {\cal A}_n(\lambda,\tilde{\lambda},\eta) = i (2\pi)^4\ \delta^{(4)}(p_{\alpha \dot\alpha})\
    \delta^{(8)}(q_{\alpha}^A)\ {\cal P}_n(\lambda,\tilde{\lambda},\eta) \,,
\ee
with the function $\mathcal{P}_n$ satisfying the relation
\be\label{conqbars}
\bar{q}_{A\, \dot \alpha} \, \mathcal{P}_n(\lambda,\tilde{\lambda},\eta) = 0\,.
\ee
Expanding $\mathcal{P}_n$ in powers of $\eta$'s and taking into account the fact that ${\cal
A}_n(\lambda,\tilde{\lambda},\eta)$ should be an $SU(4)$ singlet, we find that $\mathcal{P}_n$ is
given by
\be\label{P-sum}
\mathcal{P}_n = \mathcal{P}_n^{(0)} + \mathcal{P}_n^{(4)} + \mathcal{P}_n^{(8)} + \ldots +
\mathcal{P}_n^{(4n-16)}\,,
\ee
with $\mathcal{P}_n^{(4k)}(\lambda,\tilde{\lambda},\eta)$ being $SU(4)$ invariant homogenous
polynomials in $\eta$'s of degree $4k$.

We recall that each super-wavefunction $\Phi(p_i,\eta_i)$ carries helicity $+1$, so the total
helicity of the super-amplitude ${\cal A}_n(\lambda,\tilde{\lambda},\eta)$ equals $n$. Since each
$\eta_i^A$ has helicity $+1/2$ and the Grassmann delta function $\delta^{(8)}(q_{\alpha}^A)$ is
itself of degree 8 in the $\eta$ variables (but has vanishing helicity), the super-polynomial
$\mathcal{P}_n$ \p{P-sum} carries total helicity $n-4$. Then, the first term in the expansion
\p{P-sum}, $\mathcal{P}_n^{(0)}$, describes the MHV scattering amplitudes with  helicity $n-4$, the
second term $\mathcal{P}_n^{(4)}$ describes NMHV  scattering amplitudes with helicity $n-6$ and so
on. The last term $\mathcal{P}_n^{(4n-16)}$ corresponds to $\overline{\rm MHV}$ amplitudes with
total helicity $-(n-4)$. It has degree $4n-16$ in $\eta$, which corresponds to overall degree $4n-8$
of the amplitude (including the $\delta^{(8)}(q)$ factor).

We know that each super-wavefunction contains a term $(\eta)^4$ (see Eq.~\re{fitinto}), so the
maximal possible degree of the super-amplitude $\mathcal{A}_n$ could be $4n$. The fact that the
maximal degree is actually $4n-8$ and not $4n$ follows from the duality relation \re{dual} between
amplitudes and their Fourier conjugates. Indeed, examining the Grassmann integral on the right-hand
side of \re{dual}, it is easy to see that it maps a homogenous polynomial in $\bar\eta$ of degree
$k$ into another homogenous polynomial in $\eta$ of degree $4n-k$. This implies that, since the
minimal degree of $\mathcal{A}_n$ in \p{superamplitude} is 8, its maximal degree is $4n-8$.
\footnote{An alternative explanation, valid even in the exceptional case $n=3$, follows from the
$\bar q$ supersymmetry condition in \p{SUSY}. Its generator effectively eliminates two of the
$\eta$'s, so the maximal degree obtained from $(n-2)$ remaining $\eta$'s clearly is $4n-8$.} In a
similar manner, substituting the super-amplitude in \re{dual} by its general expression
\re{superamplitude} and comparing the terms of the same degree in $\eta$ on both sides of \re{dual},
we can establish relations between the super-polynomials $\mathcal{P}_{n}^{(4k)}$ and
$\mathcal{P}_{n}^{(4n-4k-16)}$ (with $k=0,1,2,\ldots$). We shall return to these relations in a
moment.

As was already mentioned, the function $\mathcal{P}_n^{(0)}$ describes MHV scattering amplitudes.
Comparing the super-amplitude \re{superamplitude} with the well-known expression for the tree-level
MHV gluon scattering amplitudes \cite{PT,BG}, we identify the tree-level expression for
$\mathcal{P}_n^{(0)}$ as
\begin{equation}\label{seta0}
    {\cal P}^{(0)}_{n;0} = \lr{\vev{1\, 2}\vev{2\, 3}\ldots\vev{n\, 1}}^{-1}\,.
\end{equation}
Together with \re{superamplitude}, this leads to Nair's description \cite{Nair:1988bq} of the
$n$-particle MHV tree-level super-amplitude
\begin{eqnarray}
  {\cal A}^{\rm MHV}_{n;0}(\lambda,\tilde\lambda,\eta) &=& i (2\pi)^4\frac{ \delta^{(4)}
  (\sum_{i=1}^n\ \lambda_{i}^{\alpha}\,
    \tilde\lambda_{i}^{\dot\alpha})\
    \delta^{(8)} (\sum_{i=1}^n\ \lambda_{i}^{\alpha}\, \eta^A_i)}
    {\vev{1\, 2}\vev{2\, 3}\ldots\vev{n\, 1}} \nonumber\\
  &\equiv& i (2\pi)^4\frac{ \delta^{(4)}(p)\
    \delta^{(8)} (q)}{\vev{1\, 2}\vev{2\, 3}\ldots\vev{n\, 1}}\,, \label{concorr}
\end{eqnarray}
where in the second line we have used a shorthand notation for the momentum and super-charge
conservation delta functions.

Let us now apply the relation \re{dual} to obtain the tree-level expression for
$\mathcal{P}_{n;0}^{(4n-16)}$. We insert \re{concorr} into the right-hand side of \re{dual}, replace
the variables, $\lambda\leftrightarrows\tilde\lambda$ and $\eta\to\bar\eta$, and use the integral
representation for the Grassmann delta function,
\begin{equation}\label{witgras}
    \delta^{(8)}(\sum_i\ \tilde\lambda_{i}^{\dot\alpha}\, \bar\eta_{A\, i})
    = \int d^{8}\omega\, \exp\big[{\omega^A_{\dot\alpha}\, \sum_i\, \tilde\lambda_{i}^{\dot\alpha}\,
    \bar\eta_{A\, i}}\big]\,,
\end{equation}
with $d^8\omega = \prod_{A=1}^4 \prod_{\dot\alpha=1,2} d\omega^A_{\dot\alpha}$, to obtain
\begin{align}\label{PP}
\delta^{(8)}(q_{\alpha}^A)\ \mathcal{P}_{n;0}^{(4n-16)}(\lambda,\tilde{\lambda},\eta)=
\lr{[12][23]\ldots [n1]}^{-1}\int d^{8}\omega \prod_{i=1}^n
\delta^{(4)}\lr{\eta_i^A-\tilde\lambda_i^{\dot\alpha}\omega^A_{\dot\alpha}}\,,
\end{align}
It is easy to verify that the product of delta functions on the right-hand side of \re{PP} is
proportional to $\delta^{(8)}(q_{\alpha}^A)$, since we have $q^{A\,\alpha}=\sum_i \lambda_i^\alpha
\eta_i^A = \sum_i \lambda_i^\alpha \tilde\lambda_i^{\dot\alpha}\omega^A_{\dot\alpha}
=p^{\dot\alpha\alpha} \omega^A_{\dot\alpha} =0$, by virtue of the presence of $\delta^{(4)}(p)$. Then, to determine
the polynomial $\mathcal{P}_{n;0}^{(4n-16)}$ we can integrate both sides of \re{PP} over, e.g.,
$\eta_1$ and $\eta_2$. This is done by using the  decomposition of the two-component spinor
$q_{\alpha}^A$ in the basis of the linearly independent spinors $\lambda_{1\,\alpha}$ and
$\lambda_{2\,\alpha}$,
\begin{equation}\label{decompinba}
    q_{\alpha}^A = \frac{\vev{2\, q^A}}{\vev{21}}\, \lambda_{1\,\alpha} +
    \frac{\vev{1\, q^A}}{\vev{12}}\, \lambda_{2\,\alpha}\,,
\end{equation}
and the subsequent factorization
\begin{equation}\label{foldeco}
    \delta^{(8)}(q_{\alpha}^A) = \vev{12}^4\ \delta^{(4)}\big(\eta_1^A + \frac{1}{\vev{21}} \sum_{i=3}^n \vev{2i}\eta^a_i\big)\ \delta^{(4)}\big(\eta_2^A + \frac{1}{\vev{12}} \sum_{i=3}^n \vev{1i}\eta^a_i\big)\,.
\end{equation}
The result is
\begin{align}\label{PP1}
 \mathcal{P}_{n;0}^{(4n-16)}(\lambda,\tilde{\lambda},\eta)= \lr{\vev{12}^4[12][23]\ldots [n1]}^{-1}\int
d^{8}\omega \prod_{i=3}^n
\delta^{(4)}\lr{\eta_i^A-\tilde\lambda_i^{\dot\alpha}\omega^A_{\dot\alpha}}\,.
\end{align}
In the following subsections, we will consider this relation in the special cases $n=4,5$. We will
also explain how  to treat the exceptional case $n=3$.

\subsection{Tree-level super-amplitudes for $n=4, \,5$ }

For $n=4$ the expansion \re{P-sum} involves only one term, $\mathcal{P}_4 = \mathcal{P}_{4}^{(0)}$.
This matches the fact that all non-vanishing four-particle scattering amplitudes are MHV-like. In
addition, for $n=4$ the relation \re{PP1} should be consistent with \re{seta0}. Indeed, calculating
the integral on the right-hand side of \re{PP1} by decomposing $\omega^A_{\dot\alpha}$ in the basis
of $\tilde\lambda_3$ and  $\tilde\lambda_4$ (compare to \p{decompinba}),  and making use of the
spinor identity $\vev{12}[23] = -\vev{14}[43]$ (valid for $n=4$) and similar identities obtained by
cyclic shifts of the labels, we find
\begin{align}
 \mathcal{P}_{4;0}^{(0)} = \lr{\vev{12}^4[12][23][34][41]}^{-1}[34]^4
 =\lr{\vev{12}\vev{23}\vev{34}\vev{41}}^{-1}\,,
\end{align}
in agreement with \re{seta0}.

For $n=5$ all non-vanishing amplitudes are either MHV-like, or $\rm \widebar{MHV}$-like. As a
consequence, the expansion \re{P-sum} involves two terms, $\mathcal{P}_5 =
\mathcal{P}_{5}^{(0)}+\mathcal{P}_{5}^{(4)}$. As before, $\mathcal{P}_{5}^{(0)}$ describes the
five-particle MHV amplitudes and it is given at tree level by \re{seta0}. The function
$\mathcal{P}_{5}^{(4)}$ describes the five-particle $\rm \widebar{MHV}$ amplitudes. To find its
tree-level expression, we apply \re{PP1}
\begin{align}
 \mathcal{P}_{5;0}^{(4)}(\lambda,\tilde{\lambda},\eta)= \lr{\vev{12}^4[12][23]\ldots [51]}^{-1}\int
d^{8}\omega \, \prod_{i=3}^5
\delta^{(4)}\lr{\eta_i^A-\tilde\lambda_i^{\dot\alpha}\omega^A_{\dot\alpha}}
\end{align}
and perform the $\int d^{8}\omega$ integral as in the previous case, with the result
\begin{equation}\label{p504}
    \mathcal{P}_{5;0}^{(4)}(\lambda,\tilde{\lambda},\eta)= \lr{\vev{12}^4[12][23]\ldots
    [51]}^{-1}\ \delta^{(4)}\lr{\eta_3 [45] + \eta_4 [53] +\eta_5 [34] }\,.
\end{equation}
This case is interesting because it is the simplest example of an NMHV amplitude. The argument of
the delta function in \p{p504} satisfies the condition for $\bar q$-supersymmetry \p{conqbars}, as
can be seen using the generator \p{q} and the cyclic identity for $\tilde\lambda_{3,4,5}$.

\subsection{Tree-level super-amplitudes for $n=3$ }\label{trlen3}

For $n=3$, the momentum conservation $\sum_{i=1}^3 p_i^\mu=0$ prohibits the existence of the
three-particle scattering amplitudes with real on-shell Minkowski momenta  $p_i^2=0$. However,
on-shell three-particle amplitudes can be defined if one relaxes the reality condition for the
on-shell momenta $p_i^\mu$, or changes the signature of the space time to $(++--)$. Later in the
paper, we shall follow the first route.

Similarly to \re{lambda}, the complex-valued on-shell momenta can be expressed in terms of spinors.
The only difference is that the spinors $\lambda$ and $\tilde\lambda$ are now independent complex
variables. For $i\neq j \neq k$, the condition $p_i = - (p_j+p_k)$ leads to the relation $p_i^2 =
\vev{jk} [kj] = 0$ which has two solutions, $\vev{jk} \neq 0$, $[jk]=0$ or  $\vev{jk}=0$, $[jk]\neq
0$. In the first case, we project both sides of the spinor version of the momentum conservation
condition, $\sum_{i=1}^3 \lambda_i^\alpha\tilde\lambda_i^{\dot\alpha} = 0$, with the spinor
$\tilde\lambda_k$ and obtain that the chiral spinors are proportional to each other,
\be\label{sol1}
\lambda_i^\alpha\, [ik] + \lambda_j^\alpha\, [j k] =0\,,\qquad \vev{i j} =0\,.
\ee
Analogously, for the second solution we have
\be\label{sol2}
\tilde\lambda_i^{\dot\alpha}\, \vev{ik} + \tilde\lambda_j^{\dot\alpha}\, \vev{j k} =0\,,\qquad [ij]
=0\,.
\ee
The choice of the solutions \re{sol1} or \re{sol2} we need to make is determined by whether we wish
to describe MHV or $\rm \widebar{MHV}$ tree-level amplitudes. Consider, for example, the general
expression for an $n-$particle tree-level MHV amplitude \re{concorr} and restrict it to the case
$n=3$:
\begin{equation}\label{MHV3}
    {\cal A}^{\rm MHV}_{3;0}(\lambda,\tilde\lambda,\eta) = i (2\pi)^4\delta^{(4)}(\sum_{i=1}^3\
    \lambda_{i}^{\alpha}\,
    \tilde\lambda_{i}^{\dot\alpha})\frac{
    \delta^{(8)} (\sum_{i=1}^3\ \lambda_{i}^{\alpha}\, \eta^A_i)}{\vev{1\, 2}\vev{2\, 3}\vev{3\, 1}}\,.
\end{equation}
As was already mentioned, it only exists for complex momenta. We observe that ${\cal A}^{\rm
MHV}_{3;0}(\lambda,\tilde\lambda,\eta)$ is well defined only for the kinematical configuration
\re{sol2}. Similarly, the presence of $[ij]$ in the denominator of the $n=3$ $\rm \widebar{MHV}$
amplitude (in its anti-holomorphic form) requires to make the choice \p{sol1}.

We recall that for generic $n$ the MHV amplitudes are described by the first term
$\mathcal{P}_n^{(0)}$ in the expansion \re{P-sum}, while the holomorphic description of the  $\rm
\widebar{MHV}$ amplitudes is given by the last term. However, in the exceptional case $n=3$ this
last term would have to involve a `polynomial' $\mathcal{P}_3^{(-4)}$ of negative degree. The reason
for this contradiction is that in the generic case we have always assumed that the solution of the
condition for $q$-supersymmetry  necessarily involves the factor $\delta^{(8)}(q_{\alpha}^A)$. In
fact, this is not the case for $n=3$. To see it, let us return to the relation \re{PP} and evaluate
its right-hand side for $n=3$,
\begin{align}\label{exceptdelta}
\frac{\delta^{(4)}\lr{\eta_1[23]+\eta_2[31]+\eta_3[12]}}{[12][23][31]}\,.
\end{align}
This expression has degree $4$ in $\eta$, in accord with the general formula $4n-8$ for an  $\rm
\widebar{MHV}$ amplitude and hence cannot contain the prefactor $\delta^{(8)}(q_{\alpha}^A)$.
Nevertheless, this amplitude still satisfies the condition for $q$-supersymmetry. Indeed, we can use
\p{sol1} to rewrite the generator $q^A$ in the form
\begin{equation}\label{thisspec}
    q^{A}_{\alpha} = \lambda_{1\,\alpha}\eta^A_1 + \lambda_{2\,\alpha}\eta^A_2 +\lambda_{3\,\alpha}\eta^A_3
    = \frac{\lambda_{1\,\alpha}}{[23]} \left(\eta_1 [23] + \eta_2 [31] + \eta_3 [12] \right)\,,
\end{equation}
after which it becomes clear that it annihilates the delta function in \p{exceptdelta}.

Combining this result with the condition for momentum conservation, we arrive at the exceptional
form of the $n=3$ tree-level ${\rm \widebar{MHV}}$ super-amplitude~\cite{A-H,Brandhuber:2008pf}
\be\label{bar-MHV3}
  {\cal A}^{\rm \widebar{MHV}}_{3;0}(\lambda,\tilde\lambda,\eta) = i (2\pi)^4\delta^{(4)}\big(\sum_{i=1}^3\
    \lambda_{i}^{\alpha}\,
    \tilde\lambda_{i}^{\dot\alpha}\big)
  \frac{\delta^{(4)}\lr{\eta_1[23]+\eta_2[31]+\eta_3[12]}}{[12][23][31]}\,.
\ee
In distinction with \re{MHV3}, it has degree $4$ in $\eta$ and it is well defined for the
kinematical configuration \re{sol1} only.

Since the two super-amplitudes, Eqs.~\re{MHV3} and \re{bar-MHV3}, are defined for different
kinematical configurations, Eqs.~\re{sol2} and \re{sol1}, respectively, they cannot be combined into
a single $n=3$ super-amplitude. Later, in sections \ref{sect-unitarity-MHV} and
\ref{section-unitarity-NMHV}, we will make use of the super-amplitudes ${\cal A}^{\rm {MHV}}_{3;0}$
and ${\cal A}^{\rm \widebar{MHV}}_{3;0}$ to calculate $n-$particle super-amplitudes at tree level
and at one loop using unitarity-based methods. We will illustrate the techniques on the much-studied
case of MHV amplitudes and then go on to obtain all NMHV amplitudes in the superspace form. This
extends the known case of NMHV gluon amplitudes \cite{Cachazo:2004dr,Bern:2004ky,Bern:2004bt} to
NMHV amplitudes with all possible external particles (see also \cite{Risager:2005ke} for some NMHV
amplitudes involving gluinos and scalars).

\section{Generalized unitarity for $\mathcal{N}=4$ super-amplitudes}
\label{sect-unitarity-MHV}

In the previous section we showed that all $n-$particle color-ordered scattering
amplitudes in $\mathcal{N}=4$ SYM theory can be combined into a super-amplitude
$\mathcal{A}_n(\lambda,\tilde\lambda,\eta)$. At tree level, the MHV super-amplitudes have a
particularly simple form \re{concorr}. In this section we describe an approach to calculating
one-loop corrections to the super-amplitudes. It is based on the unitary cut technique developed in
Refs.~\cite{bcf1} and it allows us to express the one-loop corrections to
$\mathcal{A}_n(\lambda,\tilde\lambda,\eta)$ as a linear combination of scalar box integrals whose
coefficients are rational functions of spinors $\lambda$ and $\bar\lambda$ and polynomials in the
odd variables $\eta$. Most importantly, we shall argue that these coefficients have a new symmetry,
dual superconformal symmetry.

\subsection{Quadruple cuts for amplitudes}

To begin with, we summarize the properties of one-loop (planar color-ordered) scattering amplitudes
$A_{n;1}$. It is known that in $\mathcal{N}=4$ SYM theory these amplitudes can be decomposed over
the basis of scalar box integrals with rational coefficients \cite{bddk94,bddk94-2}
\be\label{amplitude-decomposition}
{A}_{n;1} = \sum (c^{\rm 4m} I^{\rm 4m} + c^{\rm 3m} I^{\rm 3m} + c^{\rm 2mh} I^{\rm 2mh} + c^{\rm
2me}I^{\rm 2me} + c^{\rm 1m} I^{\rm 1m}),
\ee
where the sum runs over all possible distributions of the individual momenta of the $n$
particles. The scalar box integrals, $I^{\rm 4m}$, $I^{\rm 3m}$, $I^{\rm 2mh}$, $I^{\rm 2me}$ and
$I^{\rm 1m}$, are defined in terms of the following  dimensionally regularized integral
\be\label{I}
I(K_1,K_2,K_3,K_4) = -i (4\pi)^{2-\epsilon} \int \frac{d^{4-2\epsilon} l}{(2\pi)^{4-2\epsilon}}
\frac{1}{l^2 (l+K_1)^2 (l+K_1+K_2)^2 (l-K_4)^2}\,.
\ee
Here the momenta $K_i$ ($i=1,2,3,4$) are given by the sums of clusters of the consecutive momenta of
$n_i$ incoming particles, with $\sum_{i=1}^4 n_i = n$. For four-mass integrals $I^{\rm 4m}$, all
four momenta have non-zero invariant masses $K_{1,2,3,4}^2\neq 0$. For three-mass integrals $I^{\rm
3m}$, one of the invariant masses vanishes, e.g., $K_1^2=0$. For two-mass integrals two of the
invariant masses vanish, e.g., $K_1^2=K_2^2=0$ for $I^{\rm 2mh}$ and $K_1^2=K_3^2=0$ for $I^{\rm
2me}$. For one-mass integrals $I^{\rm 1m}$ only one invariant mass is different from zero, e.g.,
$K_4^2 \neq 0$ and $K_1^2=K_2^2=K_3^2=0$.

\begin{figure}
\psfrag{p}[cc][cc]{$K_1$} \psfrag{r}[cc][cc]{$K_2$} \psfrag{q}[cc][cc]{$K_3$}
\psfrag{s}[cc][cc]{$K_4$} \psfrag{l1}[cc][cc]{$l_1$} \psfrag{l2}[cc][cc]{$l_2$}
\psfrag{l3}[cc][cc]{$l_3$} \psfrag{l4}[cc][cc]{$l_4$} \centerline{{\epsfysize5cm
\epsfbox{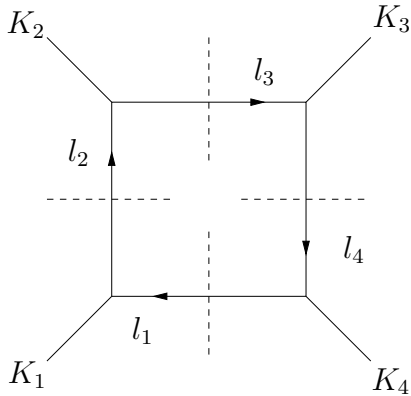}}} \caption[]{\small Quadruple cut of the scalar box integral
$I(K_1,K_2,K_3,K_4)$. The four
  cut conditions $l_i^2=0$ and momentum conservation at each corner
  leave precisely two solutions for the $l_i$.} \label{Fig:4ptcut}
\end{figure}

The dependence on the helicities of the incoming particles is carried by the coefficients $c$ and,
therefore, the problem of calculating ${A}_{n;1}$ is reduced to determining these coefficients. In
the unitary-based technique, the $c$'s are computed by comparing the analytical properties of both
sides of relation \re{amplitude-decomposition}, viewed as functions of the Mandelstam kinematical
invariants. This can be done most effectively in the generalized unitarity approach \cite{bcf1},
which makes use of the fact that each scalar box integral entering
\re{amplitude-decomposition} can be uniquely specified by their leading singularities. The latter
are obtained by cutting all four scalar propagators in \re{I} as illustrated in Fig.
\ref{Fig:4ptcut}. Since the scattering amplitudes in $\mathcal{N}=4$ SYM are
cut-reconstructible~\cite{bddk94,bddk94-2}, the cuts can be evaluated in four-dimensions.
Furthermore, the four conditions imposed by putting the four cut propagators on-shell are sufficient
to reduce the loop integration to a discrete sum over the two solutions $\mathcal{S}_\pm$ to the
on-shell and momentum conservation conditions
\be\label{cutcond}
\mathcal{S}_\pm : \qquad l_i^2=0\,,\qquad l_i^\mu+K_i^\mu=l_{i+1}^\mu\,,\qquad (i=1,2,3,4)\,,
\ee
with the periodicity condition $i+4\equiv i$. The explicit solutions are given in \cite{bcf1}, but
we do not need them for our purposes. The important point is that the four cuts completely localize
the momentum integrals. As we show in Section~3.3, the same happens to the Grassmann loop
integration in the super-amplitude with four cuts.

As discussed in detail in \cite{bcf1}, each four-particle cut selects only one term in the sum in
the right-hand side of \re{amplitude-decomposition} and, therefore, allows one to express the
corresponding coefficient $c$ in terms of the quadruple cut of the color-ordered amplitude $A_{n;1}$.
The result of the cutting procedure is that the coefficient $c$ is given by the product of four
tree-level amplitudes $A_{n_i+2;\,0}$, resulting from the cuts and evaluated on the localized
kinematical configurations \re{cutcond}
\be\label{c-4-cut}
c(n_i) = \frac12 \sum_{\mathcal{S}_\pm\!,\, J} m_J \,{A_{n_1+2;\,0} A_{n_2+2;\,0} A_{n_3+2;\,0}
A_{n_4+2;\,0}}\,,
\ee
where the sum runs over the two kinematical configurations $\mathcal{S}_\pm$ verifying the relations
\re{cutcond}, and over the possible spins $J$ of the internal particles, with $m_J$ being the number
of such particles. The values of the positive integers $n_i$ (the number of incoming particles whose
total momentum is $K_i$) determine the type of box integral in \re{c-4-cut}. In particular, for
$n_{1,2,3,4}\ge 2$ we have $K_{1,2,3,4}^2\neq 0$ and the relation \re{c-4-cut} defines the
coefficient $c^{\rm 4m}$. Similarly, for $n_1=1$ and $n_{2,3,4}\ge 2$ the corresponding coefficient
is $c^{\rm 3m}$ and so on.

\subsection{Quadruple cuts for super-amplitudes}

Let us now extend the discussion in the previous subsection to the super-amplitude $\mathcal{A}_{n;1}$. We
recall that the amplitudes $A_{n;1}$, Eq.~\re{amplitude-decomposition}, appear as coefficients in
the expansion of the super-amplitude $\mathcal{A}_{n;1}$ in powers of $\eta$'s. This allows us to
write
\be\label{super-amplitude-decomposition}
\mathcal{A}_{n;1} = i (2\pi)^4\ \delta^{(4)}(p_{\alpha \dot\alpha})\sum (\mathcal{C}^{\rm 4m} I^{\rm
4m} + \mathcal{C}^{\rm 3m} I^{\rm 3m} + \mathcal{C}^{\rm 2mh} I^{\rm 2mh} + \mathcal{C}^{\rm
2me}I^{\rm 2me} + \mathcal{C}^{\rm 1m} I^{\rm 1m})\,,
\ee
where the scalar box integrals are the same as in \re{amplitude-decomposition} and the
super-coefficients $\mathcal{C}$ now depend on $\eta$'s and include, in particular, the coefficients
$c$ entering \re{amplitude-decomposition}.

Since the relations \re{amplitude-decomposition} and \re{super-amplitude-decomposition} involve the
same scalar box integrals, particular quadruple cuts applied to the right-hand sides of the two
relations will pick out the same type of contributions, e.g., to the coefficients of the four-mass,
three-mass, etc. integrals. Just as in \re{c-4-cut}, this allows us to express the
super-coefficients $\mathcal{C}$ in terms of quadruple cut of the super-amplitude
$\mathcal{A}_{n;1}$. To write down the leading singularity of $\mathcal{A}_{n;1}$ in terms of
tree-level super-amplitudes $\mathcal{A}_{n_i+2;0}$ we use the single-particle completeness
condition
\begin{align} \label{Phi-Phi}
\int d^4\eta\, \ket{\Phi(p,\eta)} \bra{\Phi(-p,\eta)} & = \ket{G^+(p)} \bra{G^-(-p)}+\ket{G^-(p)}
\bra{G^+(-p)}
\\ \notag
& +\ket{\Gamma_A(p)} \bra{\bar\Gamma^A(-p)}+\ket{\bar\Gamma^A(p)}
\bra{\Gamma_A(-p)}+\frac12\ket{S^{AB}(p)} \bra{\bar S_{AB}(-p)}\,,
\end{align}
which can be easily verified by replacing the super-wavefunctions by their expressions \re{fitinto}
and performing the Grassmann integration. Making use of the relation \re{Phi-Phi}, we observe that
the sum over all possible particles $J$ on the right-hand side of \re{c-4-cut} can be replaced, in
the super-amplitude description, by integration over the common variable $\eta$ of two adjacent
super-amplitudes sharing the same super-wavefunction $\Phi(p,\eta)$,
\be\label{C-4-cut}
\mathcal{C}(n_i) = \frac12 \sum_{\mathcal{S}_\pm} \int d^4 \eta_{l_1}  d^4 \eta_{l_2} d^4 \eta_{l_3}
d^4 \eta_{l_4} \, {\widehat{\mathcal{A}}_{n_1+2;\,0} \, \widehat{\mathcal{A}}_{n_2+2;\,0}\,
\widehat{\mathcal{A}}_{n_3+2;\,0}\, \widehat{\mathcal{A}}_{n_4+2;\,0}}\,.
\ee
Here $\widehat{\mathcal{A}}_{n_i+2;\,0}$ stands for the tree-level
superamplitude ${\mathcal{A}}_{n_i+2;\,0}=\mathcal{A}\lr{\Phi_{l_i} \Phi_1\ldots \Phi_{n_i}
\Phi_{-l_{i+1}}}$, `amputated' of its momentum delta function,
\be\label{A-hat}
{\mathcal{A}}_{n_i+2;\,0} = i (2\pi)^4 \delta^{(4)}\big(K_i +l_i-l_{i+1}\big)
\widehat{\mathcal{A}}_{n_i+2;\,0}(l_i;\{n_i\} ;-l_{i+1})\,.
\ee
The coefficients defined in \re{C-4-cut} can be classified in the same way as in the bosonic case,
by counting the vanishing invariant masses $K_i^2=0$ (or equivalently, with $n_i=1$) in a given
kinematical configuration, e.g.,
\begin{align} \notag
& \mathcal{C}^{\rm 4m\phantom{h}} = \mathcal{C}\lr{n_{1,2,3,4}\ge 2}\,, &&  \mathcal{C}^{\rm
3m\phantom{h}} = \mathcal{C}\lr{n_1=1,n_{2,3,4}\ge 2}\,,\qquad
\\ \label{CC}
& \mathcal{C}^{\rm 2mh} = \mathcal{C}\lr{n_{1,2}=1,n_{3,4}\ge 2}\,, && \mathcal{C}^{\rm 2me} =
\mathcal{C}\lr{n_{1,3}=1,n_{2,4}\ge 2}\,,\qquad
\\ \notag
& \mathcal{C}^{\rm 1m\phantom{h}} = \mathcal{C}\lr{n_{1,2,3}=1,n_{4}\ge 2}\,.
\end{align}
By construction, the coefficients $\mathcal{C}$ are $SU(4)$ invariant polynomials in the
variables $\eta$ corresponding to the external incoming particles. As we show in the next section, they
satisfy the supersymmetry relations \re{SUSY} and therefore have the following general form (for
$n\ge 4)$
\be\label{C-gen}
\mathcal{C}^{\rm m} = 
\delta^{(8)}(\sum_{i=1}^n \lambda_i \eta_i)\left[ \mathcal{P}_{n;1}^{(0),\,\rm m} +
\mathcal{P}_{n;1}^{(4),\,\rm m} + \mathcal{P}_{n;1}^{(8),\,\rm m} + \ldots +
\mathcal{P}_{n;1}^{(4n-16),\,\rm m}\right],
\ee
where `$\rm m$' labels the five different types of coefficients and $\mathcal{P}_{n;1}^{(4k),\,\rm m}$ are
homogenous polynomials of degree $4k$ in the $\eta$'s.

\subsection{Grassmann integration}\label{gmlo}

The most straightforward way to obtain a super-amplitude  \re{C-4-cut}is by sewing together four
super-amplitudes \re{A-hat} corresponds to the four-mass case. According to \re{C-gen}, each
amplitude on the right-hand side of \re{C-4-cut} involves at least two external legs (in addition to
the two internal), $n_i+2\ge 4$. So, here we only need super-amplitudes of the conventional type
\re{superamplitude},
\be\label{A-hat1}
\widehat{\mathcal{A}}_{n_i+2;\,0}(l_i;\{n_i\} ;-l_{i+1}) =
\delta^{(8)}\big(\lambda_{l_i}\eta_{l_i}-\lambda_{l_{i+1}}\eta_{l_{i+1}}+\sum_{j\in \{n_i\}}
\lambda_j \eta_j\big)\ \mathcal{P}_{n_i+2;0}\,,
\ee
where the summation index $j$ runs over $n_i$ incoming particles inside the cluster with total
momentum $K_i$. Here $\mathcal{P}_{n_i+2;0}$ (the label $0$ means it is part of a tree-level
amplitude) is a polynomial in $\eta$ of maximal degree $4(n_i+2)-16=4n_i-8$ of the general form
\re{P-sum}. Substituting \re{A-hat1} into \re{C-4-cut}, we obtain a representation for the four-mass
coefficient $\mathcal{C}^{\rm 4m}$ in the form of a four-fold Grassmann integral containing four
Grassmann delta functions. It is easy to see that the sum of the arguments of the four delta
functions is $\sum_{i=1}^n\lambda_i \eta_i$, i.e. it only depends on the odd variables of the
external particles. This allows us to convert one of the delta functions into
$\delta^{(8)}(\sum_{i=1}^n\lambda_i \eta_i)$, in agreement with \re{C-gen}. Then, we can use two of
the remaining three delta functions to perform the integration over $\eta_{l_i}$, with the help of
the identity
\be\label{identity}
\delta^{(8)}\big(\lambda_{l_i}\eta_{l_i}-\lambda_{l_{i+1}}\eta_{l_{i+1}}+\sum_{j\in \{n_i\}}
\lambda_j \eta_j\big) = \vev{l_i\,l_{i+1}}^4 \delta^{(4)}\big(\eta_{l_i}+\sum_{j\in \{n_i\}}
\frac{\vev{l_{i+1} j}}{\vev{l_{i+1} l_i}}\eta_j \big) \delta^{(4)}\big(\eta_{l_{i+1}}-\sum_{j\in
\{n_i\}} \frac{\vev{l_i j}}{\vev{l_i l_{i+1}}}\eta_j \big)\,.
\ee
This leads to the following expression for the four-mass box coefficients $\mathcal{C}^{\rm 4m}$:
\begin{align}\label{C4m-susy}
\mathcal{C}^{\rm 4m} &= \delta^{(8)}(\sum_{i=1}^n \lambda_i \eta_i)\,\mathcal{P}_{n;1}^{\rm
4m}(\lambda,\tilde\lambda,\eta)\,,
\end{align}
where $\mathcal{P}_{n;1}^{\rm 4m}$ is given by a  product of polynomials $\mathcal{P}_{n_i+2;0}$
evaluated for the special on-shell kinematical configurations $(l_i,\eta_i)$ determined by the quadruple super-cut,
\begin{align}\label{P-4m}
\mathcal{P}_{n;1}^{\rm 4m}&=  \frac12
\sum_{\mathcal{S}^\pm}\mathcal{P}_{n_1+2;0}\mathcal{P}_{n_2+2;0}\mathcal{P}_{n_3+2;0}\mathcal{P}_{n_4+2;0}\,\vev{l_2
l_3}^4\vev{l_3 l_4}^4 \vev{l_4 l_1}^4
\\ \notag
&\times \delta^{(4)}\bigg(\sum_{j\in\{n_2\}} \eta_j
\frac{\vev{jl_2}}{\vev{l_3l_2}}+\sum_{j\in\{n_3\}} \eta_j \frac{\vev{jl_4}}{\vev{l_3l_4}} \bigg)
\delta^{(4)}\bigg(\sum_{j\in\{n_3\}} \eta_j \frac{\vev{jl_3}}{\vev{l_4l_3}}+\sum_{j\in\{n_4\}}
\eta_j \frac{\vev{jl_1}}{\vev{l_4l_1}} \bigg)\,.
\end{align}
We shall return to this relation in Section \ref{DSCSy}, where we will demonstrate that
$\mathcal{P}_n^{\rm 4m}$ has the remarkable property of dual superconformal covariance.

Let us now compute the degree of the polynomial \re{P-4m}. Since each polynomial
$\mathcal{P}_{n_i+2;0}$ has minimal degree 0 (corresponding to an MHV amplitude) and maximum degree
$4n_i-8$ (corresponding to an $\widebar{\rm MHV}$ amplitude), and the two delta functions have total
degree $8$, we find that the degree of $\mathcal{P}_{n;1}^{\rm 4m}$ ranges from 8 to $4n-24$, thus
\re{C4m-susy} becomes
\be\label{P-4m-gen}
\mathcal{C}^{\rm 4m} =   \delta^{(8)}(\sum_{i=1}^n \lambda_i \eta_i)\,\left[
\mathcal{P}_{n;1}^{(8),\,{\rm 4m}} + \ldots + \mathcal{P}_{n;1}^{(4n-24),\,{\rm 4m}} \right]\,.
\ee
Comparing this relation to the general expression \re{C-gen}, we observe that the first two and the
last two terms in \re{C-gen} are absent in the expansion \re{P-4m-gen},
$\mathcal{P}_{n;1}^{(4k),\,{\rm 4m}} = 0$ for $k=0,2$ and for $k=n-5,n-4$. We recall that the degree
of homogeneity of the polynomial $\mathcal{P}_{n}^{(4k)}$ is in one-to-one correspondence with the
helicity configuration of the underlying scattering amplitudes. Then, the relation \re{P-4m-gen}
implies that the four-mass box terms in \re{super-amplitude-decomposition} do not contribute to the
MHV and NMHV super-amplitudes (as well as their Fourier conjugates $\widebar{\rm MHV}$ and
$\widebar{\rm NMHV}$). To describe the MHV and NMHV super-amplitudes, we have to consider box
integrals (and their coefficients) with at least one vanishing invariant mass.

In the generalized cut approach, the presence of integrals with one or more vanishing invariant
masses implies that we have to include the exceptional three-particle super-amplitudes  in
\re{A-hat}. As discussed in Sect.~3.3, there are two types of such amplitudes. The first is the
MHV three-particle amplitude \re{MHV3}, leading to the corresponding amputated super-amplitude given
by the expression \re{A-hat1} for, e.g., $n_i=1$,
\be\label{MHV-amp}
\widehat{\mathcal{A}}^{\ \rm {MHV}}_{3;\,0}(l_i;1;-l_{i+1}) =\frac{
\delta^{(8)}\big(\lambda_{l_i}\eta_{l_i}-\lambda_{l_{i+1}}\eta_{l_{i+1}}+  \lambda_1
\eta_1\big)}{\vev{l_1 1} \vev{1 l_2} \vev{l_2 l_1}}\,.
\ee
The second is the three-particle $\widebar{\rm MHV}$ super-amplitude ${\cal A}^{\rm \widebar{MHV}}_{3;0}$ defined in
\re{bar-MHV3}. A special feature of the latter is that the corresponding amputated super-amplitude
\re{A-hat},
\be\label{barMHV-amp}
\widehat{\mathcal{A}}^{\ \rm \widebar{MHV}}_{3;\,0}(l_i;1;-l_{i+1})
=\frac{\delta^{(4)}\lr{\eta_{l_1}[1 l_2]+\eta_1[l_2l_1]+\eta_{l_2}[l_11]}}{[l_1 1][1 l_2][l_2 l_1]}\,,
\ee
has degree of homogeneity in $\eta$ equal to four (to be compared with \re{A-hat1} or \re{MHV-amp},
whose minimal degree is 8). This modifies the counting of $\eta$'s in \re{C-4-cut}, as we show
below. Armed with these three-point super-amplitudes, in addition to the conventional \re{A-hat1},
we can calculate the remaining coefficients $\mathcal{C}$ in \re{CC}.

First, let us examine the relation \re{C-4-cut} for the three-mass box coefficients
$\mathcal{C}^{\rm 3m}$. Choosing the massless leg to be $n_1=1$, we have to replace $
\widehat{\mathcal{A}}_{n_1+2;\,0}$ in \re{C-4-cut} by the sum of two amputated super-amplitudes
\re{barMHV-amp} and \re{MHV-amp}. As before, the four-fold Grassmann integral over $\eta_{l_i}$ can
easily be done with the help of the Grassmann delta functions and, in close analogy with
\re{C4m-susy}, the resulting expression for $\mathcal{C}^{\rm 3m}$ takes the form
\begin{align}\label{C3m-exp}
\mathcal{C}^{\rm 3m} &= \delta^{(8)}(\sum_{i=1}^n \lambda_i \eta_i)\,\left[
\mathcal{P}_{n;1}^{(4),\,{\rm 3m}} + \ldots + \mathcal{P}_{n;1}^{(4n-20),\,{\rm 3m}}\right]\,.
\end{align}
Here the first term has the lowest possible degree 12, so it can only be obtained by using the
three-particle $\widebar{\rm MHV}$ vertex \re{barMHV-amp}. This term contributes to the NMHV
super-amplitude (see Section \ref{section-unitarity-NMHV} for the detailed calculation). Similarly,
the last term in \re{C3m-exp} has the maximal allowed degree $4n-12$, so it originates from the
three-particle MHV vertex \re{MHV-amp} and contributes to the $\widebar{\rm NMHV}$ super-amplitude.
All the intermediate terms in \re{C3m-exp} can get two types of contributions, with the MHV or the
$\widebar{\rm NMHV}$ three-particle vertex.

\begin{figure}
\psfrag{p1}[cc][cc]{$p_1$} \psfrag{p2}[cc][cc]{$p_2$} \centerline{{\epsfysize3cm
\epsfbox{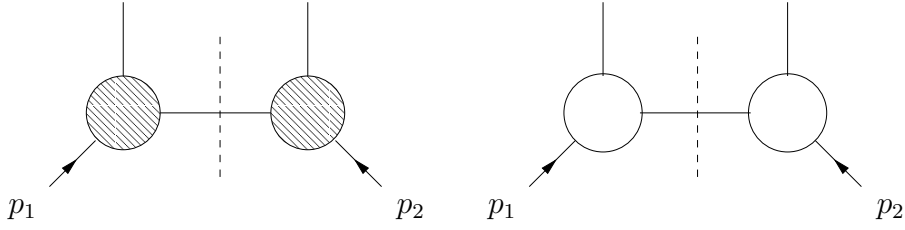}}} \caption[]{\small
  Two adjacent three-point $\bMHV$ or two adjacent three-point MHV vertices. In either
  case the on-shell momentum conservation conditions imply that
  $(p_1+p_2)^2=0$ so the configuration does not exist for general
  kinematics.} \label{Fig:3ptadj}
\end{figure}

For the two-mass box coefficients in \re{CC}, $\mathcal{C}^{\rm 2mh}$ and $\mathcal{C}^{\rm 2me}$,
two of the super-amplitudes on the right-hand side of \re{C-4-cut} should be replaced with the
three-particle super-amplitudes \re{barMHV-amp} and/or \re{MHV-amp}. We recall that the amplitudes
$\widehat{\mathcal{A}}^{\ \rm \widebar{MHV}}_{3;\,0}$ and $\widehat{\mathcal{A}}^{\ \rm
{MHV}}_{3;\,0}$ are defined for two different kinematical configurations, Eqs.~\re{sol1} and
\re{sol2}, respectively. If two such sub-amplitudes of the same type are adjacent to each other, say
$\widehat{\mathcal{A}}^{\ \rm {MHV}}_{3;\,0}(l_1,1,-l_2)\widehat{\mathcal{A}}^{\ \rm
{MHV}}_{3;\,0}(l_2,2,-l_3)$, then the kinematical constraints \re{sol2} for each of them lead to the
proportionality of the spinor variables $\tilde\lambda_1 \propto \tilde\lambda_2 $ with the
corollary $[12]=0$. However, this cannot be satisfied for general kinematics,
$(p_1+p_2)^2=\vev{12}[21]\neq 0$. Therefore, the two three-point ${\rm MHV}$ vertices must be placed
at opposite corners of the cut box, and it is possible to have at most two such vertices. The same
constraints apply to the three-point $\overline{\rm MHV}$ vertices. These statements are summarised
in Fig. \ref{Fig:3ptadj}.
For the two-mass-hard box coefficient $\mathcal{C}^{\rm 2mh}$, the corresponding box diagram
contains three-point ${\rm MHV}$ and $\overline{\rm MHV}$  vertices adjacent to each other (see
Fig.~Fig:2mh). We substitute $\widehat{\mathcal{A}}_{n_1+2;\,0} \to \widehat{\mathcal{A}}^{\ \rm
{MHV}}_{3;\,0}$ and $\widehat{\mathcal{A}}_{n_2+2;\,0} \to \widehat{\mathcal{A}}^{\ \rm
\widebar{MHV}}_{3;\,0}$ in \re{C-4-cut} and use the expression \re{A-hat1} for two remaining
sub-amplitudes to find that the integral over $\eta_{l_i}$ is again localized by the Grassmann delta
functions at the vertices, leading to
\begin{align}\label{C2mh-exp}
\mathcal{C}^{\rm 2mh} &= \delta^{(8)}(\sum_{i=1}^n \lambda_i \eta_i)\,\left[
\mathcal{P}_{n;1}^{(4),\,{\rm 2mh}} + \ldots + \mathcal{P}_{n;1}^{(4n-20),\,{\rm 2mh}}\right]\,.
\end{align}
Thus, the two-mass coefficient contribute to all super-amplitudes except the MHV and $\widebar{\rm MHV}$
ones.

For the two-mass-easy box coefficient $\mathcal{C}^{\rm 2me}$, the corresponding box diagram involves
two three-particle ${\rm MHV}$ and/or $\overline{\rm MHV}$ vertices situated at two opposite corners of the box.
The minimal (or maximal) degree in $\eta$ is achieved when both  three-particle vertices are $\widebar{\rm
MHV}$ (or MHV). Performing the calculation of \re{C-4-cut} we find
\begin{align}\label{C2me-exp}
\mathcal{C}^{\rm 2me} &= \delta^{(8)}(\sum_{i=1}^n \lambda_i \eta_i)\,\left[
\mathcal{P}_{n;1}^{(0),\,{\rm 2me}} + \ldots + \mathcal{P}_{n;1}^{(4n-16),\,{\rm 2me}}\right]\,.
\end{align}

Finally, the one-mass box coefficient $\mathcal{C}^{\rm 1m}$ corresponds to a box diagram in which
three of the vertices are three-particle ${\rm MHV}$ and/or $\overline{\rm MHV}$ ones. We recall that two
three-particle vertices of the same type can not be adjacent. After some algebra, we find from \re{C-4-cut} that
\begin{align}\label{C1m-exp}
\mathcal{C}^{\rm 1m} &= \delta^{(8)}(\sum_{i=1}^n \lambda_i
\eta_i)\,\left[\mathcal{P}_{n;1}^{(0),\,{\rm 1m}} + \ldots + \mathcal{P}_{n;1}^{(4n-16),\,{\rm 1m}}
\right]\,.
%
\end{align}
We conclude that $\mathcal{C}^{\rm 2me}$ and  $\mathcal{C}^{\rm 1m}$ contribute to all
super-amplitudes and these are the only two coefficients that contribute to the MHV and
$\widebar{\rm MHV}$ super-amplitudes. In the next subsection, as an illustration of the general
scheme developed here, we compute the corresponding contributions in the MHV case,
$\mathcal{P}_{n;1}^{(0),\,{\rm 2me}}$ and $\mathcal{P}_{n;1}^{(0),\,{\rm 1m}}$.

\subsection{One-loop MHV super-amplitude}

\begin{figure}
\psfrag{p}[cc][cc]{$p_1$} \psfrag{a}[cc][cc]{$p_2$} \psfrag{b}[cc][cc]{$p_{s-1}$}
\psfrag{q}[cc][cc]{$p_s$} \psfrag{e}[cc][cc]{$p_{s+1}$} \psfrag{f}[cc][cc]{$p_n$}
\psfrag{p}[cc][cc]{$1$} \psfrag{a}[cc][cc]{$2$} \psfrag{b}[cc][cc]{${s-1}$} \psfrag{q}[cc][cc]{$s$}
\psfrag{e}[lc][cc]{${s+1}$} \psfrag{f}[rc][cc]{$n$}
\centerline{{\epsfysize5cm \epsfbox{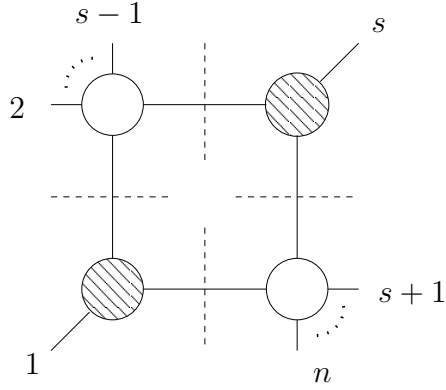}}} \caption[]{\small
  The only allowed configuration contributing to the one-loop MHV super-amplitude. It
  corresponds to a cut two-mass easy integral in the general case. If
  $s=3$ or $s=n-1$ then it is a one-mass integral and if $s=3=n-1$ then it
  is a massless box.} \label{Fig:MHV4cut}
\end{figure}

The MHV super-amplitude receives contributions from the terms on the right-hand side of
\re{C2me-exp} and \re{C1m-exp} with lowest degree in $\eta$'s. Such terms come from the diagram with
two three-point $\overline{\rm MHV}$ vertices at opposite corners of the cut box. The lowest
possible degree for each of the other two vertices is 8 (corresponding to MHV with any number of
points). In this case the Grassmann degree of the resulting total super-amplitude is $4+4+8+8-16=8$,
which is precisely what is needed for an MHV super-amplitude. This configuration corresponds in
general to a two-mass-easy coefficient $\mathcal{C}^{\rm 2me}$ and is illustrated in Fig.
\ref{Fig:MHV4cut}. In the special case where one of the MHV vertices is a three-particle vertex, the
same diagram defines a one-mass coefficient $\mathcal{C}^{\rm 1m}$.~\footnote{For a four-particle
amplitude, both MHV vertices can be three-particle vertices in which case the configuration
corresponds to a massless box coefficient.}

Let us compute the contribution of the diagram shown in Fig.~\ref{Fig:MHV4cut} following the scheme
described in the previous subsection. We start with the general expression \re{C-4-cut} and substitute
the super-amplitudes,
\begin{align}\label{repl}
& \mathcal{\widehat A}_{n_1+2;0}\to {\mathcal{\widehat A}}^{\ \rm \widebar{MHV}} (l_1;1;-l_{2})\,,
&&\mathcal{\widehat A}_{n_2+2;0}\to {\mathcal{\widehat A}}^{\ \rm {MHV}}
(l_2,2,\ldots,s-1,-l_{3})\,,
\\[2mm] \notag
& \mathcal{\widehat A}_{n_3+2;0}\to {\mathcal{\widehat A}}^{\ \rm \widebar{MHV}} (l_3;s;-l_{4})\,,
&&\mathcal{\widehat A}_{n_4+2;0}\to {\mathcal{\widehat A}}^{\ \rm {MHV}}
(l_4,s+1,\ldots,n,-l_{1})\,,
\end{align}
where the three-particle ${\rm \widebar{MHV}}$ super-amplitude is given by \re{bar-MHV3} and the tree-level
$n-$particle MHV super-amplitude is defined in \re{A-hat1} and \re{seta0}. In this way, we obtain
\begin{align} \label{c2meMHV}
\mathcal{C}_{1,2,s,s+1} =  \frac12 \sum_{\mathcal{S}_\pm}\int \prod_{i=1}^4 d\eta_{l_i}
&\frac{\delta^{(4)}(\eta_1 [l_2 l_1] + \eta_{l_2}[l_1 1] +\eta_{l_1}[1
    l_2])}{[1 l_2][l_2 l_1][l_1 1]}
\frac{\delta^{(8)}(\lambda_{l_2}\eta_{l_2} + \sum_2^{s-1} \lambda_i \eta_i
  - \lambda_{l_3} \eta_{l_3})}{\l<l_2 2\r>\ldots\l<s-1\,\,l_3\r>\l<l_3
  l_2\r>} \notag \\
&\frac{\delta^{(4)}(\eta_{l_3}[s l_4] + \eta_s[l_4 l_3] + \eta_{l_4}[l_3
    s])}{[l_3 s][s l_4][l_4 l_3]}\frac{\delta^{(8)}(\lambda_{l_4}
  \eta_{l_4} + \sum_{s+1}^n \lambda_i \eta_i -
  \lambda_{l_1}\eta_{l_1})}{\l<l_4\,\,s+1\r>\ldots\l<n l_1\r>\l<l_1 l_4\r>},
\end{align}
where the sum goes over the two kinematical configurations  \re{cutcond} with $K_1=p_1$, $K_2=\sum_2^{s-1}
p_i$, $K_3=p_s$ and $K_4=\sum_{s+1}^n p_i$. The four labels of $\mathcal{C}_{1,2,s,s+1}$ indicate the first (clockwise) particle in each cluster.

It is straightforward to compute the four Grassmann integrals on the right-hand side of \re{c2meMHV}
with the help of the identity \re{identity}. Here we prefer to present a shortcut, which makes
efficient use of the symmetry of the problem. We remark that the coefficient \p{c2meMHV} corresponds
to a particular term in a superamplitude \re{super-amplitude-decomposition}, and therefore is expected to
be invariant under $q$--supersymmetry. This is easy to verify by inserting the generator  $q=
\sum_{i=1}^n \lambda_i\eta_i$ (see \re{q}) under the integrals in  \p{c2meMHV}, and then distributing it over
the four delta functions (we recall the property \p{thisspec} of the three-particle vertices).
Consequently, the result of the integration in \p{c2meMHV} must be proportional to
$\delta^{(8)}(\sum_{i=1}^n \lambda_i\eta_i)$. On the other hand, the expression \p{c2meMHV} is of
degree 8 in the $\eta$'s and, therefore, its entire $\eta$ dependence is contained in this
$\delta^{(8)}(\sum_{i=1}^n \lambda_i\eta_i)$. In order to detect its presence, it is sufficient to
keep any subset of at least two external $\eta$'s, while setting the rest to zero. For instance, we
can choose to set $\eta_1=\eta_s=0$, as well as $\eta_i=0$ with $i=2,\ldots,s-1$ if $s\neq 3$ (or
alternatively,  with $i=s+1,\ldots,n$ if $s\neq n-1$). Then the second delta function in \p{c2meMHV}
factorizes into
\begin{equation}\label{delfacto}
    \delta^{(8)}(\lambda_{l_2}\eta_{l_2} - \lambda_{l_3}\eta_{l_3}) =
    \vev{l_2 l_3}^4\ \delta^{(4)}(\eta_{l_2}) \  \delta^{(4)}(\eta_{l_3})\,,
\end{equation}
after which the first and the third delta functions become simply $[1l_2]^4\
\delta^{(4)}(\eta_{l_1})$ and $[l_3 s]^4\ \delta^{(4)}(\eta_{l_4})$, respectively. This allows us to
trivially do all four integrals, leaving just $\delta^{(8)}(\sum_{s+1}^{n} \lambda_i \eta_i )$
which, after restoring the missing external $\eta$'s, becomes $\delta^{(8)}(\sum_{i=1}^n
\lambda_i\eta_i)$. Finally, collecting the various spinor factors, we find
\be
\mathcal{C}_{1,2,s,s+1} = \frac{\delta^{(8)}(\sum_{1}^{n} \lambda_i \eta_i)}{\vev{12}\vev{23}\ldots
\vev{n1}}\Delta_{1,2,s,s+1}\,,
\ee
where the scalar factor $\Delta_{1,2,s,s+1}$ is given by
\be
\Delta_{1,2,s,s+1} =  \vev{s-1\, s}\vev{s \, s+1}\vev{n\, 1}\vev{1\,2}\times \frac12
\sum_{\mathcal{S}_\pm}\frac{[1| l_2 l_3 |s]^2}{\vev{s-1|l_3 l_4 |s+1}\vev{2|l_2 l_1|n}}\,.
\ee
Here the $l_i$ satisfy the on-shell conditions \re{cutcond} and the standard conventions for
contraction of spinors and light-like vectors were used, e.g., $\vev{i|l_j l_k|l} = \vev{ij}[jk]\vev{kl}$ and $[i|l_j l_k|l] = [ij]\vev{jk}[kl]$.
Then, we take into account the relations $l_1=l_2-p_1$ and $l_4=l_3+p_s$ to simplify $\vev{2|l_2
l_1|n}=\vev{2|p_1l_1|n} = \vev{21}[1l_1]\vev{l_1n}$ and similarly for the second factor in the
denominator. After some algebra we find
\begin{align} \notag
\Delta_{1,2,s,s+1} & = \frac12 \sum_{\mathcal{S}_\pm} \frac{\vev{s-1\, s}\vev{n\,1}}{\vev{s-1\,
l_3}\vev{n\,l_1}} [1\, l_1] [l_3\, s] \vev{l_1\, l_3}^2
\\ \notag
& = -\frac12 \sum_{\mathcal{S}_\pm} \vev{l_2l_1} [l_1 l_3] \vev{l_3 l_4} [l_4 l_2]
\\ \label{Delta-2me}
& = \frac14 \sum_{\mathcal{S}_\pm} \left[(l_2-l_3)^2 (l_1-l_4)^2-(l_1-l_3)^2(l_2-l_4)^2 \right].
\end{align}
Here in the second line we used the kinematical relations \re{sol1} between the chiral spinors
$\lambda_{l_1}, \lambda_1$ and $\lambda_{l_3}, \lambda_s$, imposed by the three-particle vertices $\rm
\widebar{MHV}$ in \re{repl}. We observe that the invariant masses $(l_i-l_j)^2$ are uniquely fixed
by the kinematical invariants $K_{1,2,3,4}^2$ and, therefore, the sum in \re{Delta-2me} can be
evaluated without using the explicit form of the solutions for $l_i^\mu$
\begin{align} \notag
\Delta_{1,2,s,s+1} & = \frac12 \mbox{$\left[\lr{\sum_{2}^{s-1}p_i}^2 \lr{\sum_{1}^s p_i}^2 -
\lr{\sum_{1}^{s-1}p_i}^2 \lr{\sum^{s}_{2}p_i}^2 \right]$}
\\
& = \frac12 \left[x_{2\,s}^2 x_{1\, s+1}^2-x_{1\,s}^2 x_{2\, s+1}^2 \right]\,,
\end{align}
where in the second relation we switched to the dual variables $p_i = x_i - x_{i+1}$ (see \re{dualx}
below).

In terms of the dual variables, the momenta $K_i$ entering the four vertices of the box diagram shown
in Fig.~\ref{Fig:4ptcut} are given by
\begin{align}\label{K-2me}
K_1=x_{1\,2},\qquad K_2=x_{2\,s},\qquad K_3=x_{s\, s+1},\qquad K_4=x_{s+1\, 1}\,.
\end{align}
By the definition \re{CC}, the two-mass easy coefficients $\mathcal{C}_{1,2,s,s+1}$ have
$K_2^2=x_{2s}^2\neq 0$ and $K_4^2=x_{1\, s+1}^2\neq 0$. This leads to the condition  $4\le s \le
n-2$. For $s=3$ and $s=n-1$ one of the MHV vertices in the box diagram shown in
Fig.~\ref{Fig:MHV4cut} reduces to a three-particle MHV vertex and it defines the one-mass coefficient
$\mathcal{C}^{\rm 1m}$. This allows us to combine the contributions of the two-mass easy and
one-mass coefficients to the one-loop MHV superamplitude into
\be\label{1-loop-MHV}
\mathcal{A}_{n;1}^{\rm MHV} = i (2\pi)^4\ \delta^{(4)}(p_{\alpha \dot\alpha})
\frac{\delta^{(8)}(\sum_{1}^{n} \lambda_i \eta_i)}{\vev{12}\vev{23}\ldots \vev{n1}} \left[
\sum_{s=3}^{n-1} I_{1,2,s,s+1}\Delta_{1,2,s,s+1} + \text{cyclic} \right]\,,
\ee
where `cyclic' stands for the terms needed to restore the symmetry of the super-amplitude under
cyclic shifts of the indices of the incoming particles. Also, $I_{1,2,s,s+1}$ denotes the scalar box integral
\re{I} evaluated for the kinematical configuration \re{K-2me},
\be
I_{1,2,s,s+1}\equiv I(K_1,K_2,K_3,K_4) = \frac{F_{1,2,s,{s+1}}}{\Delta_{1,2,s,s+1}}\,.
\ee
Here in the second relation, $F_{1,2,s,{s+1}}$ is a dimensionless translation invariant function of
the dual coordinates $x_1,x_2,x_s,x_{s+1}$. It contains infrared divergences which appear in the
dimensional regularization scheme with $D=4-2\epsilon$ as poles in $\epsilon$. The explicit form of
this function can be found in Appendix~A.

Finally, comparing \re{1-loop-MHV} with the tree-level expression for the MHV super-amplitude
\re{concorr}, we conclude that the one-loop corrections to $\mathcal{A}_{n}^{\rm MHV}$ appear as a
scalar factor given by the sum of dimensionless scalar box functions
\be\label{dual-A}
\mathcal{A}_{n;1}^{\rm MHV} = \mathcal{A}_{n;0}^{\rm MHV}\times  \left[ \sum_{s=3}^{n-1}
F_{1,2,s,{s+1}} + \text{cyclic} \right].
\ee
Since this property is a consequence of supersymmetry, it holds to all loops. Most remarkably, the
MHV superamplitude $\mathcal{A}_{n}^{\rm MHV}$ was conjectured~\cite{am1,dks,bht} to be dual to the
expectation value of Wilson loop $W_n$ evaluated along a closed contour composed of light-like
momenta of incoming particles $p_i$ (with $i=1,\ldots,n$)
\be\label{dual-W}
\mathcal{A}_{n}^{\rm MHV} /W_n = \mathcal{A}_{n;0}^{\rm MHV} \left[ 1 + O(\epsilon)\right]\,.
\ee
The one-loop corrections to $W_n=1 + g^2 N c_\Gamma W_{n;1} + O(g^4)$ can be expressed (up to an
additive constant correction) in terms of two-mass easy and one-mass scalar box integrals
\be\label{dual-W1}
W_{n;1} = \frac12 \sum_{r=1}^{n}\sum_{s=r+2}^{r+n-2} F_{r,r+1,s,s+1}\,,
\ee
with indices defined modulo $n$. It is easy to see that the relations \re{dual-A} and \re{dual-W}
indeed coincide to one loop. The duality relation \re{dual-W} has been
verified~\cite{dhks3,bdkrsvv,dhks4} by an explicit two-loop calculation for $n=6$ and was shown to
hold at strong coupling within AdS/CFT correspondence~\cite{am1,am2}.

\section{NMHV super-amplitudes}
\label{section-unitarity-NMHV}

In this section, we apply the generalized unitarity method to compute the one-loop corrections to
the next-to-MHV (NMHV) super-amplitudes. As a byproduct, we obtain a new and very compact
representation for the tree-level NMHV super-amplitudes. \footnote{This form of the NMHV tree
amplitudes was first conjectured in \cite{dhks5} and compared to the NMHV gluon tree amplitude form
\cite{Bern:2004ky}.}

We would like to mention that some six-point NMHV amplitudes were computed in \cite{Huang} using
two-particle cuts and supersymmetric vertices. However, the Grassmann calculation was only carried
out explicitly for certain types of external particles. It is straightforward to extend the
calculation of \cite{Huang} to include arbitrary external particles. In complete analogy with the
MHV case in section \ref{sect-unitarity-MHV}, one can factor out a Grassmann delta function
containing the dependence on the external particle super-momenta, and the loop algebra is done as in
the bosonic case.  We do not present the calculation here because the generalised cut technique is
more efficient when going to a higher number of external points. The reason is that in the bosonic
two-particle cut calculation the box integrals appear only after employing integral reduction
techniques. This complication does not arise when using generalised cuts because the latter allow us
to single out one box integral coefficient at a time, as was already explained.

We recall that the general one-loop super-amplitude is given by the linear combination
\re{super-amplitude-decomposition} of the possible scalar box integrals with the appropriate
coefficients $\mathcal{C}$ of the form \re{C-gen}. According to \re{P-4m-gen}, the four-mass
coefficient first appears in the NNMHV amplitudes, so we
have%
\footnote{There are always at least six external particles for an NMHV amplitude and hence the
zero-mass box four-particle configuration does not appear.}
\be
A_{n;1}^{\rm NMHV} = \sum (\mathcal{C}^{\rm 3m} I^{\rm 3m} + \mathcal{C}^{\rm 2mh} I^{\rm 2mh} +
\mathcal{C}^{\rm 2me}I^{\rm 2me} + \mathcal{C}^{\rm 1m} I^{\rm 1m})\big|_{\rm \scriptscriptstyle
NMHV}\,.
\ee
Here the sum runs over all possible distributions of the individual momenta and the superscript
`NMHV' indicates that inside the coefficients $\mathcal{C}$ \re{C-gen} we retain only the
contribution of degree 12 in the $\eta$'s.

\subsection{Three-mass and two-mass-hard coefficients}\label{3mcose}

\begin{figure}
\psfrag{=}[cc][cc]{$\Longleftrightarrow$}\psfrag{dots}[cc][cc]{$\bf\ldots$}
\psfrag{p}[cc][cc]{$r$} \psfrag{a}[cc][cc]{$r+1$} \psfrag{b}[cc][cc]{$s-1$} \psfrag{c}[cc][tc]{$s$}
\psfrag{d}[cc][cc]{$t-1$} \psfrag{e}[rc][cc]{$t$} \psfrag{f}[rc][cc]{$r-1$}
\psfrag{p1}[cc][cc]{$p_1$} \psfrag{p2}[cc][cc]{$p_2$} \psfrag{p3}[cc][cc]{$p_3$}
\psfrag{p4}[cc][bc]{$\vdots $} \psfrag{p5}[cc][cc]{$p_{n-1}$} \psfrag{p6}[cc][cc]{$p_n$}
\centerline{{\epsfysize5.5cm \epsfbox{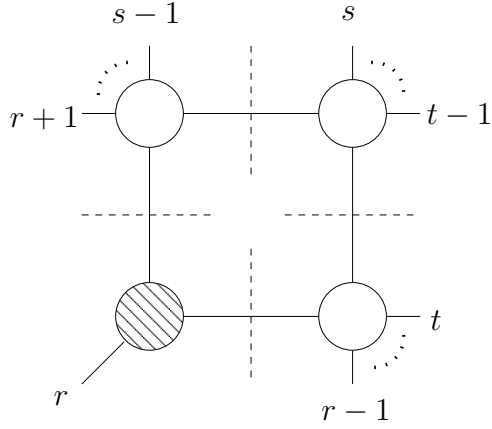}}} \caption[]{\small The configuration contributing
to the three-mass coefficient $\mathcal{C}^{\rm 3m}_{r,r+1,s,t}$. The empty vertices are MHV
super-amplitudes and the shaded vertex is a three-particle $\overline{\rm MHV}$ super-amplitude.
} \label{Fig:3mass}
\end{figure}

We begin by calculating the three-mass coefficients $\mathcal{C}^{\rm
3m}\big|_{\rm\scriptscriptstyle NMHV}$. The corresponding cut-box diagram is shown in Fig.
\ref{Fig:3mass}. It contains one three-point vertex and three vertices with four legs or more. To
produce a contribution of degree 12 in $\eta$ (i.e. NMHV), the former  should
be a three-particle $\overline{\rm MHV}$ vertex \re{barMHV-amp} while the latter are generic
MHV vertices.

According to \re{C-4-cut}, gluing these vertices together corresponds to performing the following
Grassmann integrations,
\begin{align}
\mathcal{C}^{\rm 3m}_{r,r+1,s,t} = \int \prod_{i=1}^4 d\eta_{l_i} &
  \frac{\delta^{(4)}(\eta_r [l_2 l_1] +
  \eta_{l_2}[l_1 r] + \eta_{l_1}[r l_2])}{[r l_2][l_2 l_1] [l_1 r]}
  \frac{\delta^{(8)}(\eta_{l_2} \lambda_{l_2}
+ \sum_{r+1}^{s-1} \eta_i \lambda_i - \eta_{l_3} \lambda_{l_3})}{\l<
  l_2 r+1 \r>
  \l<r+1 \, r+2\r>\ldots \l<s-1\, l_3\r> \l<l_3 l_2\r>}\notag\\
\times&\frac{\delta^{(8)}(\eta_{l_3} \lambda_{l_3} + \sum_s^{t-1} \eta_i
  \lambda_i - \eta_{l_4} \lambda_{l_4})}{\l<l_3 s\r>
  \l<s\, s+1\r>\ldots\l<t-1\, l_4\r>\l<l_4 l_3\r>} \frac{\delta^{(8)}(\eta_{l_4}
  \lambda_{l_4} + \sum_{t}^{r-1} \eta_i \lambda_i - \eta_{l_1}
  \lambda_{l_1})}{\l<l_4 t\r>\l<t\, t+1\r>\ldots\l<r-1\, l_1 \r>\l<l_1 l_4\r>}\,.
\label{c3mass}
\end{align}
Here the sums in the argument of the three $\delta^{(8)}$ functions run over the incoming particles
entering the three MHV vertices. If the upper limit of a sum is actually lower than the lower limit,
the sum is to be understood in the cyclic sense, i.e. $\sum_s^t = \sum_s^n + \sum_1^t$. The momenta $l_i$ satisfy the on-shell conditions as usual. Only one of the two solutions to the cut conditions contributes to the coefficient; the other solution would require a three-point MHV vertex (instead of $\bMHV$).

As before, the Grassmann delta functions localize the integrals over $\eta_{l_i}$ in \re{c3mass}. To
simplify the calculation, we rewrite the third delta function in \re{c3mass} by adding to its
argument the sum of the arguments of the other two $\delta^{(8)}$ functions, thus obtaining
$\delta^{(8)}(\eta_{l_2} \lambda_{l_2} + \sum_{r+1}^{r-1} \eta_i \lambda_i - \eta_{l_1}
\lambda_{l_1})$. Then we eliminate $\eta_{l_1}$ and $\eta_{l_2}$ from the argument of the latter,
using the first delta function in \re{c3mass}:
\be
\eta_{l_2} \lambda_{l_2} - \eta_{l_1} \lambda_{l_1}  + \sum_{r+1}^{r-1} \eta_i \lambda_i =-
\eta_{l_1} \lambda_{l_1} - \lr{\eta_r \frac{[l_1l_2]}{[r l_1]}+\eta_{l_1}\frac{[l_2 r]}{[r
l_1]}}\lambda_{l_2}  + \sum_{r+1}^{r-1} \eta_i \lambda_i = \sum_{1}^n \eta_i \lambda_i\,,
\ee
where in the last relation we used the kinematical constraints $\lambda_{l_2} [l_2 r] =
\lambda_{l_1} [l_1 r]$ and $\lambda_{l_2} [l_1 l_2] = \lambda_r [l_1 r]$, coming from the
three-particle $\widebar{\rm MHV}$ vertex \re{sol1} (see also Fig.~\ref{Fig:3mass}). Thus, we have
obtained the expected super-momentum conservation delta function.  Finally, we use the second and
the fourth delta functions in \re{c3mass} to perform the integrations, leading to
\begin{align}\label{integrated}
\mathcal{C}^{\rm 3m}_{r,r+1,s,t} &=  \frac{[r l_1]^4 }{[l_3 l_4]^4 D}
 \delta^{(4)}\Bigl( \sum_{t}^{r-1} \eta_i \vev{i|l_4 l_3 |l_1}+
 \sum_{r}^{s-1}\eta_i \vev{i|l_3 l_4 |l_1} \Bigr) {\delta^{(8)}\Bigl(\sum_{1}^n
\eta_i \lambda_i\Bigr)}\,,
\end{align}
where $D$ represents all the denominator factors in (\ref{c3mass}) and the identity
$\vev{i|l_j l_k |l_1} = \vev{i l_j} [l_j l_k]\vev{l_k l_1}$ was used. The factors of the $\eta_i$'s can be
further simplified as
\be\label{ex1}
\vev{i|l_4 l_3|l_1} = \vev{i|(l_4-l_3)(l_3-l_1)|l_1} =
\vev{i|\Big(\sum_{s}^{t-1}p_j\Big)\Big(\sum_{r}^{s-1}p_k\Big) |l_1} = - \frac{[r l_2]}{[l_1
l_2]}\vev{i|x_{ts}x_{sr} |r}\,.
\ee
Here in the last relation we expressed the on-shell momenta in terms of the dual variables $p_i=x_i-x_{i+1}$
and used the relation between the chiral spinors at the three-particle $\widebar{\rm MHV}$ vertex
\re{sol1}. Treating the factors  $\vev{i|l_3 l_4|l_1}$ similarly, we can rewrite \re{integrated} in the form
\begin{align}\label{C3m-D}
\mathcal{C}^{\rm 3m}_{r,r+1,s,t} &= \frac{[r l_1]^4 [r l_2]^4}{[l_3 l_4]^4[l_1 l_2]^4 D}\
\delta^{(4)}\bigl(\Xi_{rst}\bigr)\ {\delta^{(8)}\Bigl(\sum_{1}^n \eta_i \lambda_i\Bigr)} \,,
\end{align}
where $\Xi_{rst}$ is defined as
\be\label{iniforxi}
\Xi_{rst} =\sum_{t}^{r-1} \eta_i \vev{i|x_{ts}x_{sr} |r} +
 \sum_{r}^{s-1}\eta_i \vev{i|x_{st}x_{tr} |r}\,.
\ee
Finally, we replace $D$ in \re{C3m-D} by the product of all denominator factors in (\ref{c3mass})
and obtain, after some algebra,
\begin{align}\label{3mcoeff}
\mathcal{C}^{\rm 3m}_{r,r+1,s,t} &= \Delta_{r,
r+1,s,t}\frac{c_{rst}\delta^{(4)}\bigl(\Xi_{rst}\bigr)}{\prod_1^n \vev{i \, i+1}}
 {\delta^{(8)}\bigl(q\bigr)}\,,
\end{align}
where $q=\sum_1^n \lambda_i \eta_i$ and the notation was introduced for
\begin{align}\notag
& \Delta_{r,r+1,s,t} = -\frac{[l_1 r][l_2 r] \vev{l_3 r}\vev{l_4 r}[l_3 l_4]}{  [l_1 l_2]}
=\tfrac12\left[(l_1-l_3)^2 (l_2-l_4)^2-(l_1-l_4)^2(l_2-l_3)^2 \right]\,,\qquad
\\[3mm]
& c_{rst} =
-\frac{\vev{s-1\,s}\vev{t-1\,t}}{(l_3-l_4)^2\vev{r|l_3l_4|s-1}\vev{r|l_3l_4|s}\vev{r|l_4l_3|t-1}\vev{r|l_4l_3|t}}\,.
\end{align}
The expression for $c_{rst}$ can be further simplified along the same lines as in \re{ex1}. Going to
dual variables, we find
\begin{align}\notag
& \Delta_{r,r+1,s,t} = \tfrac12\left[x_{rs}^2x_{r+1t}^2-x_{rt}^2 x_{r+1 s}^2 \right]\,,\qquad
\\[3mm]
& c_{rst} = -\frac{\vev{s-1\,s}\vev{t-1\,t}}{x_{st}^2\vev{r|x_{rt}x_{ts}|s-1}\vev{r|x_{rt}x_{ts}|s}
\vev{r|x_{rs}x_{st}|t-1}\vev{r|x_{rs}x_{st}|t}}\,.
\end{align}
The factor $\Delta_{r,r+1,s,t}$ is exactly what is needed to convert the dimensionful integral
$I_{r,r+1,s,t}$ into a dimensionless function $F_{r,r+1,s,t}$ (see Appendix).
\begin{figure}
\psfrag{p}[cc][cc]{$r$} \psfrag{a}[cc][cc]{$r+1$} \psfrag{b}[cc][cc]{$s-1$} \psfrag{c}[cc][cc]{$s$}
\psfrag{x}[cc][tc]{$r-1$} \psfrag{d}[cc][cc]{$r-2$} \psfrag{f}[rc][cc]{$t$}
\centerline{{\epsfysize5.5cm \epsfbox{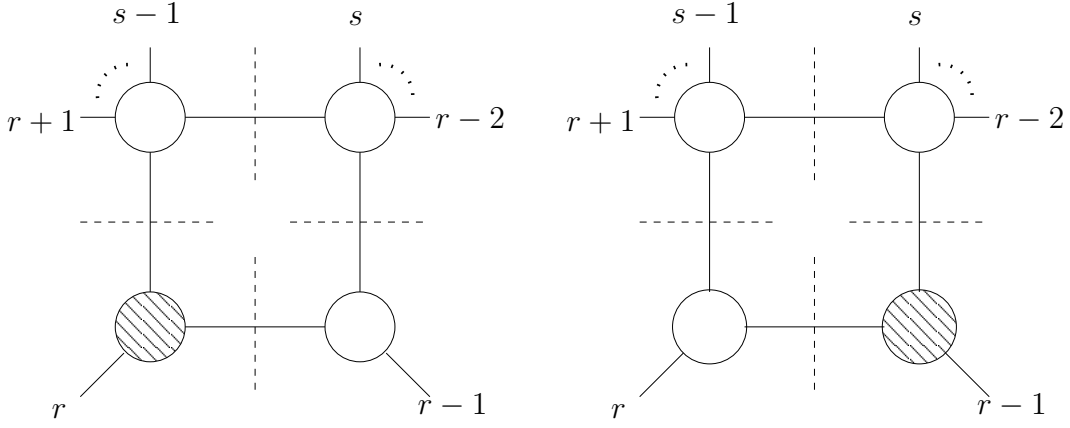}}} \caption[]{\small The two configurations
contributing to the two-mass-hard coefficient $\mathcal{C}^{\rm 2mh}_{r-1,r,r+1,s}$. They are simply
related to the three-mass coefficients $\mathcal{C}^{\rm
  3m}_{r,r+1,s,r-1}$ (left) and $\mathcal{C}^{\rm 3m}_{r-1,r,r+1,s}$ (right).} \label{Fig:2mh}
\end{figure}

To summarize, the contribution of the three-mass coefficient to the one-loop NMHV amplitude takes the form
\be\label{sumrun}
 \sum_{r, s, t} \mathcal{C}^{\rm 3m}_{r,r+1,s,t} I_{r,r+1,s,t} =
 \frac{\delta^{(8)}\bigl(q\bigr)}{\prod_1^n \vev{i \, i+1}}
 \sum_{r, s, t} R_{rst} \,F_{r,r+1,s,t}
\ee
with \footnote{In section \ref{sect-superdual} we recall that $R_{rst}$ is the three-point dual superconformal invariant introduced and studied in \cite{dhks5}.}
\be\label{Omega}
R_{rst} =  -\frac{\vev{s-1\,s}\vev{t-1\,t}\  \delta^{(4)}\bigl(\Xi_{rst}\bigr)}{x_{st}^2\vev{r|x_{rt}x_{ts}|s-1}\vev{r|x_{rt}x_{ts}|s}
\vev{r|x_{rs}x_{st}|t-1}\vev{r|x_{rs}x_{st}|t}}\,.
\ee
The sum in \p{sumrun} runs over the indices $r,s,t=1,\ldots, n$ which satisfy the relations $s-r > 2 \ \mbox{(mod
$n$)}$, $t-s>1 \ \mbox{(mod $n$)}$ and $r-t> 1 \ \mbox{(mod $n$)}$, determined by the kinematics of the
three-mass box diagram shown in Fig.~\ref{Fig:3mass}.

We notice that for $t=r-1$ (or $r+1=s-1$), the box diagram in Fig.~\ref{Fig:3mass} reduces to the
two-mass hard contribution to the NMHV amplitude. This allows us to simply adapt the above
calculation to the new case. We must remember however that for a given two-mass-hard integral $I_{r-1,r,r+1,s}$
there are two contributions, shown in Fig.~\ref{Fig:2mh}, which must be added up. Thus we conclude that
the two-mass hard box integral (with three-point vertices attached to legs $r-1$ and $r$) comes with the
coefficient
\be \label{2mhcoeff}
\mathcal{C}^{\rm 2mh}_{r-1,r,r+1,s} =  \mathcal{C}^{\rm
  3m}_{r,r+1,s,r-1}+ \mathcal{C}^{\rm 3m}_{r-1,r,r+1,s}\,,
\ee
where the indices $r,s=1,\ldots,n$ have to satisfy the condition $s-r > 2 \ \mbox{(mod $n$)}$. The
relation \re{2mhcoeff} generalises a similar relation for $n$-gluon
NMHV amplitudes obtained in \cite{Bern:2004bt}.

\subsection{Two-mass-easy and one-mass coefficients}

\begin{figure}
\psfrag{=}[cc][cc]{$\Longleftrightarrow$}\psfrag{dots}[cc][cc]{$\bf\ldots$}
\psfrag{p}[cc][cc]{$r$} \psfrag{a}[cc][cc]{$r+1$} \psfrag{b}[cc][cc]{$s-1$} \psfrag{x}[cc][tc]{$s$}
\psfrag{d}[cc][cc]{$s+1$} \psfrag{e}[lc][cc]{$s+1$} \psfrag{f}[rc][cc]{$r-1$}
\centerline{{\epsfysize5.5cm \epsfbox{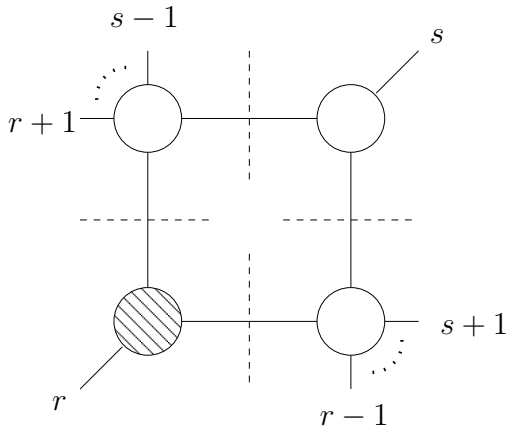}}} \caption[]{\small This configuration of vertices
vanishes for general kinematics and so does not contribute to the two-mass-easy coefficient.}
\label{Fig:2me0}
\end{figure}

For two-mass easy and one-mass integral coefficients the direct calculation from generalised cuts is
more involved because there are contributions from NMHV tree amplitudes.

\begin{figure}
\psfrag{p}[cc][cc]{$r$} \psfrag{a}[cc][bc]{$r+1$} \psfrag{b}[cc][cc]{$s-1$}
\psfrag{e}[cc][bc]{$s+1$} \psfrag{x}[cc][tc]{$s$} \psfrag{N}[cc][cc]{N} \psfrag{f}[rc][cc]{$r-1$}
\centerline{{\epsfysize5.5cm \epsfbox{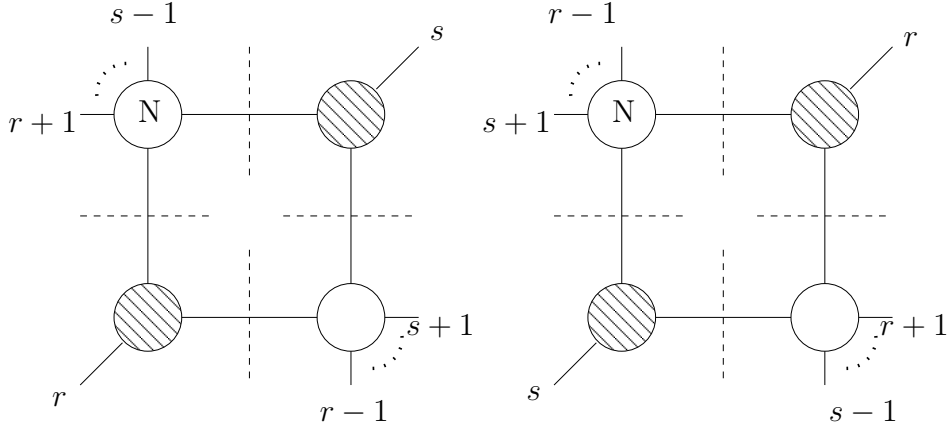}}} \caption[]{\small The non-vanishing contributions
to the two-mass-easy coefficient $\mathcal{C}^{\rm 2me}_{r,r+1,s,s+1}$. They involve NMHV tree-level
subamplitudes, indicated by the vertices with the label N.} \label{Fig:2meN}
\end{figure}

For two-mass-easy coefficients, the NMHV contribution of degree 12 in $\eta$ comes from the cut-box
diagrams shown in Figs.~\ref{Fig:2me0} and \ref{Fig:2meN}. The diagram in Fig.~\ref{Fig:2me0}
involves three-particle MHV and $\widebar{\rm MHV}$ vertices which impose the kinematical constraints
\re{sol1} and \re{sol2} on the (anti)chiral spinors at these vertices. It is
straightforward to verify that these constraints impose the relation $(l_1-l_3)^2
(l_2-l_4)^2=(l_2-l_3)^2 (l_4-l_1)^2$ which is not satisfied for general kinematics. Therefore, the
two-mass-easy coefficients only receive contribution from the diagrams shown in Fig.~\ref{Fig:2meN}.

\subsubsection{Tree-level NMHV super-amplitude}

To calculate the two-mass-easy coefficients, we need an expression for tree-level NMHV
super-amplitude. In the general expression for the $n-$particle super-amplitude, Eqs.~\re{superamplitude}
and \re{P-sum}, these amplitudes are described by the polynomial $\mathcal{P}_{n;0}^{(4)}$ and have Grassmann
degree $12$.

Let us begin with the observation that the five-point $\overline{\rm MHV}$ amplitude can be regarded
as an NMHV amplitude.  Indeed, it is defined to tree level by the polynomial \re{p504} of degree 4.
According to \re{superamplitude} and \re{p504}, the corresponding amputated super-amplitude reads%
\footnote{We remind that throughout the paper we denote the super-amplitudes stripped of the
momentum delta function by a hat.}
\begin{align}
\widehat{\mathcal{A}}^{\ \rm NMHV}_{5;0} =   \delta^{(8)}\big({\sum_1^5 \lambda_i\eta_i}\big)
 \frac{\delta^{(4)}\lr{\eta_3 [45] + \eta_4 [53] +\eta_5 [34]}}{\vev{12}^4[12][23][34][45][51]}\,.
\end{align}
A remarkable feature of this relation is that the expression on the right-hand side can be rewritten
as a product of five-point MHV super-amplitude $\widehat{\mathcal{A}}^{\ \rm MHV}_{5;0}$,
Eq.~\re{concorr}, and the coefficients $R$ defined in \re{Omega},
\be\label{A5-NMHV}
\widehat{\mathcal{A}}^{\ \rm NMHV}_{5;0} = \widehat{\mathcal{A}}^{\ \rm MHV}_{5;0} \,R_{241} =
\widehat{\mathcal{A}}^{\ \rm MHV}_{5;0}\times \lr{ \frac 15 \sum_{r=1}^5 R_{r,r+2,r+4}}\,,
\ee
with indices satisfying the periodicity condition $r+5\equiv r$. The second, manifestly cyclic
symmetric form becomes possible due to the identity for the superinvariants \cite{dhks5}
\begin{equation}\label{idfrompap}
    R_{r,r+2,s} = R_{r+2,s,r+1}\,,
\end{equation}
valid for arbitrary $n$ with the periodicity condition for indices $r+n\equiv r$. Applied to the
case $n=5$, it gives, e.g., $R_{241}= R_{413}$. From this, doing cyclic shifts $i \to i+1$, we
obtain the rest of the terms in the sum in \p{A5-NMHV}.

The formula for the five-point NMHV tree amplitude \re{A5-NMHV} is a special case of the following
general formula for the NMHV tree-level super-amplitudes first conjectured in \cite{dhks5},
\be\label{gentree}
\widehat{\mathcal{A}}^{\ \rm NMHV}_{n;0} = \widehat{\mathcal{A}}^{\ \rm MHV}_{n;0} \sum_{s,t=1}^n
R_{1st} = \widehat{\mathcal{A}}^{\ \rm MHV}_{n;0}\times \lr{ \frac 1n \sum_{r,s,t=1}^n R_{rst}}\,,
\ee
where $R_{rst}$ is given by \re{Omega} and the sum runs over the indices satisfying the relations
\be\label{indices}
s-r \ge 2 \ \mbox{(mod $n$)}\,,\qquad  t-s \ge 2 \ \mbox{(mod $n$)}\,,\qquad r-t \ge 1 \ \mbox{(mod
$n$)}\,.
\ee
It is convenient to use a diagrammatic representation of $R_{rst}$ as the cut-box diagram shown in
Fig.~\ref{Fig:3mass}. Then, the conditions \re{indices} correspond to all possible diagrams in which
two vertices adjacent to the shaded vertex have one or more legs attached to them and the remaining
vertex has two and more legs attached.

In eq.~\re{gentree}, the second, manifestly cyclic symmetric form follows from the
identity~\footnote{For $n=5$ the two identifies \p{idfrompap} and \p{genident} are equivalent. }
\be\label{genident}
\sum_{s,t=1}^n R_{1st} = \sum_{s,t=1}^n R_{nst}\,,
\ee
in which the indices satisfy the same conditions \re{indices} with $r=1$ in the left sum and $r=n$
in the right sum. For $n=5$ and $n=6$ the identity \re{genident} reads
\be\label{R=R}
R_{135} = R_{524}\,, \qquad R_{135} +R_{136} +R_{146} = R_{624} +R_{625} + R_{635}\,.
\ee
We would like to stress that these relations do not use neither the momentum conservation
$\delta^{(4)}(p)$, nor supermomenta conservation $\delta^{(8)}(q)$ and, therefore, they are
fulfilled for arbitrary number of particles. In particular, we can apply the cyclic shift of indices
$i\to i+k$ (with $k$ arbitrary) to both sides of \re{R=R} to obtain a new set of identities.

While the simple identity \p{idfrompap} is quite straightforward to prove (see \cite{dhks5}), the
new identity \p{genident} is very non-trivial. At present we do not have an analytic proof for it,
but have checked it numerically for $n=6,7$. For future use, note that if  the above identities are
valid for some number $n$ of external particles, then they are automatically valid for any $n' > n$,
provided we do not change the values of the labels and we have not use the cyclic periodicity
condition. Thus, in the case $n=5$ the identity \p{idfrompap} implies $R_{135}= R_{352}$, without
using the periodicity condition $i+5=i$. Then this identity is valid for any $n \geq 5$, but the
identity $R_{241}= R_{413}$ (obtained from the former by a cyclic shift) only applies to the case
$n=5$.

\subsubsection{Gluing tree-level NMHV super-amplitudes}

We are now ready to compute remaining two-mass easy and one-mass coefficients and, then, obtain the
complete one-loop NMHV superamplitude. In the process we will also derive the formula
(\ref{gentree}) for the NMHV tree-level super-amplitude. To do this we will proceed inductively.
That is, we know that the tree-level formula (\ref{gentree}) holds for the case $n=5$. We will
assume it holds for all $m$-particle amplitudes up to $m=n-1$. Then we will calculate the one-loop
$n$-point two-mass-easy and one-mass integral coeffiecients. Given these and the known three-mass
and two-mass-hard coefficients, Eqs.~(\ref{3mcoeff}) and (\ref{2mhcoeff}), we will know the full
$n$-point one-loop NMHV super-amplitude. Examining the infrared singularities of this
super-amplitude and using the well-known fact that the residue at infrared poles should be
proportional to the tree amplitude \cite{Bern:2004bt,Roiban:2004ix,bcf2}, we will be able to deduce
the form (\ref{gentree}) for the $n$-point tree-level NMHV super-amplitude, which will complete the
induction.

Let us perform the inductive step. We begin with the two-mass-easy coefficients, $\mathcal{C}^{\rm
2me}_{r,r+1,s,s+1}$ corresponding to the configurations shown in Fig. \ref{Fig:2meN}. Since the
second diagram in  Fig. \ref{Fig:2meN} can be obtained from the first one through substitution $r
\leftrightarrows s$, we can take $r<s$ without loss of generality. We have
\begin{align}
\mathcal{C}^{\rm 2me}_{r,r+1,s,s+1} = \int \prod_i d^4\eta_{l_i} & \frac{\delta^{(4)}(\eta_r [l_2
l_1] + \eta_{l_2}[l_1 r] + \eta_{l_1}[r l_2])}{[l_2 l_1] [l_1 r] [r l_2]}
\frac{\delta^{(8)}(\lambda_{l_4} \eta_{l_4} + \sum_{s+1}^{r-1} \lambda_i \eta_i -
\lambda_{l_1}\eta_{l_1})} {\l<l_4 \,\,s+1\r>...\l<r-1\,\,l_1\r>\l<l_1 l_4\r>} \notag
\\ \label{C-2me}
\times & \frac{\delta^{(4)}(\eta_s [l_4 l_3] + \eta_{l_4} [l_3 s] + \eta_{l_3}[s l_4])}{[l_4 l_3]
[l_3 s] [s l_4]} \,\,\widehat{\mathcal{A}}_{s-r+1;0}^{\ \rm NMHV} + (r \leftrightarrows s)\,,
\end{align}
where $\widehat{\mathcal{A}}_{s-r+1;0}^{\ \rm NMHV}$ denotes the $(s-r+1)$-particle tree-level
(amputated) NMHV super-amplitude and the other three super-amplitudes are written explicitly. As
before, the sum over two kinematical configurations $\frac12\sum_{\mathcal{S}^\pm}$ is tacitly
assumed on the right-hand side of \re{C-2me}.

We assume that the formula (\ref{gentree}) holds for the $m$-point tree-level super-amplitude for
all $m<n$. Since the tree-level NMHV super-amplitude entering \re{C-2me} has $m=s-r+1 <n$ legs, we
can use \re{gentree} to find $\widehat{\mathcal{A}}_{s-r+1;0}^{\ \rm NMHV}$. Note that, by virtue of
the identity  (\ref{genident}), there is a freedom in choosing the first label of ${R}$ in the first
relation in (\ref{gentree}). The choice of this label does not affect
$\widehat{\mathcal{A}}_{s-r+1;0}^{\ \rm NMHV}$ but it allows us to obtain different equivalent
expressions for $\mathcal{C}^{\rm 2me}_{r,r+1,s,s+1}$.

The first way singles out the cut leg $l_2$ as the first label of ${R}$ so that the NMHV tree-level
super-amplitude (\ref{gentree}) takes the form
\begin{align}
\label{NMHVfact1stform} \widehat{\mathcal{A}}_{s-r+1;0}^{\ \rm NMHV} &=  \Bigl(\prod_j
\l<j\,j+1\r>\Bigr)^{-1}\delta^{(8)}\Big(\lambda_{l_2} \eta_{l_2} + \sum_{r+1}^{s-1} \lambda_i \eta_i
- \lambda_{l_3} \eta_{l_3}\Big) \sum_{u,v}{R}_{l_2uv}\,,
\end{align}
where the indices $u$ and $v$ in the sum and $j$ in the product run over the cyclically ordered set
$\{l_2,r+1,r+2,...,s-2,s-1,l_3\}$ with the constraints $u\geq r+2$ and $u+2\leq v \leq l_3$. Also,
${R}_{l_2uv}$ is given by a general expression \re{Omega} with indices $r,s,t$ replaced with
$l_2,u,v$, respectively. Notice that ${R}_{l_2uv}$ defined in this way does not depend on
$\eta_{l_2}$. The second way of writing the NMHV tree-level factor singles out the leg $l_3$ in
which case the same super-amplitude takes the form
\begin{align}
\label{NMHVfact2ndform} \widehat{\mathcal{A}}_{s-r+1;0}^{\rm NMHV} &=  \Bigl(\prod_j \l<j
\,j+1\r>\Bigr)^{-1}\delta^{(8)}\Big(\lambda_{l_2} \eta_{l_2} + \sum_{r+1}^{s-1} \lambda_i \eta_i -
\lambda_{l_3} \eta_{l_3}\Big) \sum_{u,v}{R}_{l_3uv}\,,
\end{align}
where the indices $u$ and $v$ in the sum and $j$ in the product run over the cyclically ordered set
$\{l_3,l_2,r+1,r+2,...,s-2,s-1\}$ with the constraints $u\geq r+1$ and $u+2\leq v \leq s-1$.

It is important to note that in both forms (\ref{NMHVfact1stform}) and (\ref{NMHVfact2ndform}) the
dependence of $\widehat{\mathcal{A}}_{s-r+1;0}^{\rm NMHV}$ on $\eta_{l_2}$ and $\eta_{l_3}$ only
resides in the $\delta^{(8)}(\ldots)$ factor. This means that with either way of writing the
tree-level NMHV super-amplitude, performing the Grassmann integration in \re{C-2me} is essentially
identical to the MHV case (\ref{c2meMHV}) illustrated in the previous section. Substituting
\re{NMHVfact1stform} into \re{C-2me} and going through the same steps as in Sect.~3.4, we arrive at
\begin{align}
\mathcal{C}^{\rm 2me}_{r,r+1,s,s+1} = &\frac{[s l_4]^4 \l<l_4 l_3\r>^4 [l_3 r]^4
\delta^{(8)}\bigl(\sum_1^n \lambda_i \eta_i\bigr)   \, \sum_{u,v}{R}_{l_2uv}} {\l<l_2
\,\,r+1\r>...\l<s-1\,\,l_3\r>\l<l_3 l_2\r>
 [l_2 l_1] [l_1 1] [1 l_2] \l<l_4 \,\,s+1\r>...\l<r-1\,\,l_3\r>\l<l_3 l_2\r> [l_4 l_3] [l_3 s] [s l_4]} \notag \\
& + (r \leftrightarrows s).\label{C-2me-1}
\end{align}
We next note that from the kinematical condition \re{sol1} imposed by the three-particle
$\widebar{\rm MHV}$ vertex we have $\lambda_{l_2} = \lambda_r [r l_1]/[l_2 l_1]$. We apply this
identity to substitute for $\lambda_{l_2}$ in \re{C-2me-1} and use \re{Omega} to observe from that
the constant of proportionality cancels inside ${R}_{l_2,u,v}$ so that the label $l_2$ can simply be
replaced by $r$.

Finally, the simplification of the remaining factors in \re{C-2me-1} is identical to the MHV case and
we arrive at
\be
\mathcal{C}^{\rm 2me}_{r,r+1,s,s+1} = \sum_{u,v} \mathcal{C}^{\rm 3m}_{r,r+1,u,v} + (r
\leftrightarrows s), \label{2mecoeff}
\ee
where the indices are summed in the first term with the constraints $u\geq r+2$ and $u+2\leq v \leq
s$ and in the second term similarly but where $r$ is swapped with $s$. In the expression for the
one-loop NMHV super-amplitude \re{super-amplitude-decomposition} the coefficient \re{2mecoeff} is
accompanied by the corresponding scalar box integral $I_{r,r+1,s,s+1}$ (see Appendix A).

The relation \re{2mecoeff} generalises a similar relation for gluon NMHV amplitudes obtained in
\cite{Bern:2004bt} (see Eq.~(28) there). Here, however, it applies not only to gluon amplitudes but to
the whole NMHV super-amplitude. The relation  \re{2mecoeff} admits a simple diagrammatic
representation similar to Fig.~4 in \cite{Bern:2004bt}. We recall that three-mass coefficients
$\mathcal{C}^{\rm 3m}_{r,r+1,s,t}$ are described by the cut-box diagram shown in
Fig.~\ref{Fig:3mass}. Then, the first sum on the right-hand side of \re{2mecoeff} corresponds to
various rearrangements of legs attached to three `massive' vertices in such a way that all legs
except the leg with the label $r$ are moved in the `clockwise' direction. If we take the second form
of the NMHV tree-level factor (\ref{NMHVfact2ndform}) and repeat the same calculation, we arrive at
the same formula (\ref{2mecoeff}) with the only difference that now the indices are summed with the
constraints $u\geq r+1$ and $u+2\leq v \leq s-1$ in the first term and similarly where $r$ is
swapped with $s$ in the second. This produces another, `anti-clockwise' representation for the same
coefficient $\mathcal{C}^{\rm 2me}_{r,r+1,s,s+1}$. As was already explained, the two representations
are equivalent thanks to the identity \re{genident}. This proves the `handedness' condition
formulated in \cite{Bern:2004bt} for gluon NMHV amplitudes.

\begin{figure}
\psfrag{p}[cc][cc]{$r$} \psfrag{a}[cc][cc]{$r+1$} \psfrag{b}[cc][cc]{$r-3$}
\psfrag{e}[cc][cc]{$s+1$} \psfrag{x}[cc][tc]{$r-2$} \psfrag{N}[cc][cc]{N} \psfrag{f}[rc][cc]{$r-1$}
\psfrag{xx}[cc][cc]{$r-1$}\psfrag{yy}[cc][cc]{$r-1$}
\centerline{{\epsfysize5.5cm
\epsfbox{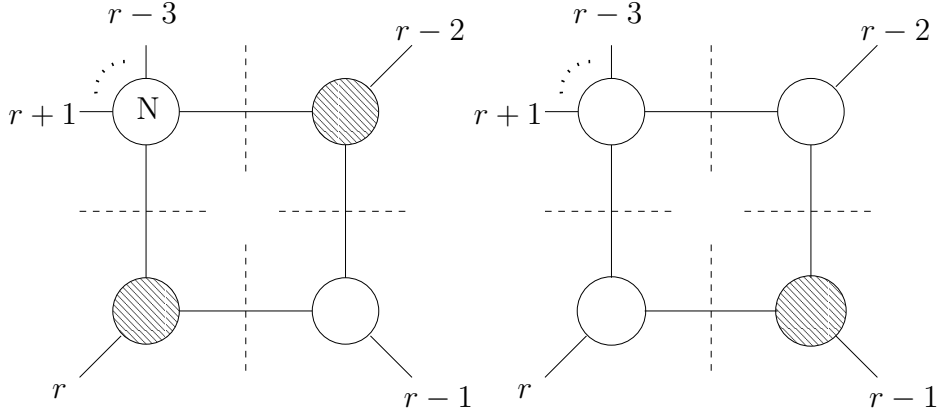}}} \caption[]{\small The two contributions to the one-mass coefficient
$\mathcal{C}^{\rm 1m}_{r-2,r-1,r,r+1}$. The first is obtained from the two-mass-easy coefficient
$\mathcal{C}^{\rm 2me}_{r,r+1,r-2,r-1}$  and the second from the three-mass coefficient
$\mathcal{C}^{\rm 3m}_{r-1,r,r+1,r-2}$ by restricting the all corners but one to be three-point
vertices.} \label{Fig:1m}
\end{figure}

The one-mass coefficients are now simple to calculate. There are two contributions which are
illustrated in Fig. \ref{Fig:1m}. The first one can be deduced from the two-mass-easy calculation as a
limiting case of the cut-box diagram shown in Fig.~\ref{Fig:2meN} for $s+1=r-1$. The second one
follows from the three-mass calculation as a limit of the cut-box diagram shown in
Fig.~\ref{Fig:3mass}. In this way we find
\be\label{C-1m}
\mathcal{C}^{\rm 1m}_{r-2,r-1,r,r+1} = \mathcal{C}^{\rm 2me}_{r,r+1,r-2,r-1} + \mathcal{C}^{\rm
  3m}_{r-1,r,r+1,r-2}\,.
\ee
This coefficient is accompanied by the scalar box integral $I_{r-2,r-1,r,r+1}$ (see Appendix A). The
relation \re{C-1m} generalises a similar relation for the gluon amplitudes found in \cite{Bern:2004bt} (see
Eq.~(33)).

To summarise, assuming the form (\ref{gentree}) for the $m$-particle tree-level NMHV super-amplitude
for $m\leq n-1$, we have computed all $\mathcal{C}-$coefficients. Their substitution into
\re{super-amplitude-decomposition} yields the complete one-loop $n$-particle NMHV super-amplitude
$\widehat{\mathcal{A}}_{n;1}^{\ \rm NMHV}$. From this we can obtain the $n$-particle tree-level
super-amplitude to complete the induction. To find the form of the tree-level amplitude
$\widehat{\mathcal{A}}_{n;0}^{\ \rm NMHV}$ it suffices to look at the infrared poles of the one-loop
amplitude which are known to have the following universal form%
\footnote{Here the parameter of the perturbative expansion is $g^2 N c_\Gamma \mu^{2\epsilon}$ with
$c_\Gamma$ given by \re{c-G}.}
\be\label{IR-poles}
\widehat{\mathcal{A}}_{n;1}^{\ \rm NMHV}\bigg|_{\epsilon\ {\rm poles}} = -\frac{ 1}{\epsilon^2}
\sum_{i=1}^n \lr{{-s_{i,i+1}}}^{-\epsilon} \times\widehat{\mathcal{A}}_{n;0}^{\ \rm NMHV}\,.
\ee
This will be done in the next subsection.

\subsection{One-loop NMHV super-amplitudes}\label{4.3}

Let us use the expressions for the $\mathcal{C}-$coefficients obtained in the previous subsection to
compute the one-loop NMHV superamplitudes for $n=6$ and $n=7$ particles.

\subsubsection{Six-point NMHV super-amplitude}

For $n=6$, due to small number of external particles, the one-loop super-amplitude receives
contributions from two-mass-hard and one-mass integrals
\be\label{A6-dec}
\widehat{\mathcal{A}}^{\ \rm NMHV}_{6;1} = \mathcal{C}^{\rm 2mh}_{1246}\, I_{1246} +
\mathcal{C}^{\rm 1m}_{1234} \,I_{1234} + \text{cyclic}\,,
\ee
where `cyclic' stands for terms obtained by cyclic shift of all indices $i\to i+1$ with the periodicity
condition $i+6\equiv i$. We apply the relations \re{2mhcoeff} and \re{C-1m} to express the
two-mass-hard and one-mass coefficients in terms of $\mathcal{C}^{3m}-$coefficients
\begin{align} \notag
\mathcal{C}^{\rm 2mh}_{1246} & = \mathcal{C}^{\rm 3m}_{1246} + \mathcal{C}^{\rm 3m}_{6124} \,,\qquad
\\[3mm]
\mathcal{C}^{\rm 1m}_{1234} &= \mathcal{C}^{\rm 3m}_{2341} +\mathcal{C}^{\rm 2me}_{1234} =
\mathcal{C}^{\rm 3m}_{2341} + \mathcal{C}^{\rm 3m}_{1246} = \mathcal{C}^{\rm 3m}_{2341} +
\mathcal{C}^{\rm 3m}_{3451}\,,
\end{align}
Here the two relations for $\mathcal{C}^{\rm 1m}_{1234}$ originate from the equivalence of the clockwise and anti-clockwise
representations for the two-mass-easy coefficient $\mathcal{C}^{\rm 2me}_{1234}$. As we will see in a
moment, the consistency condition $\mathcal{C}^{\rm 3m}_{1246}=\mathcal{C}^{\rm 3m}_{3451}$ leads to
the relation between ${R}$ coefficients which is just a special case of a general relation
\re{genident}.

 Let us replace the $\mathcal{C}^{\rm 3m}$ coefficients and the scalar box integrals by their
expressions in terms of ${R}$ and dimensionless functions $F_{r,r+1,s,t}$, Eqs.~\re{3mcoeff} and
\re{sumrun},
\begin{align}\label{C-I}
\mathcal{C}^{\rm 3m}_{r,r+1,s,t} &= \Delta_{r,r+1,s,t} {R}_{rst}\, \widehat{\mathcal{A}}^{\ \rm
MHV}_{n;0}\,,\qquad
\\ \notag
I_{r,r+1,s,t} & = \Delta_{r,r+1,s,t}^{-1} F_{r,r+1,s,t}\,.
\end{align}
Then, we combine the various terms on the right-hand side of \re{A6-dec} containing ${R}$
coefficients with the same indices to find
\begin{align}\notag
\widehat{\mathcal{A}}^{\ \rm NMHV}_{6;1} &= \widehat{\mathcal{A}}^{\ \rm MHV}_{6;0}\left[
\lr{R_{146}+R_{624}}F_{1246}+ \lr{R_{241}+R_{146}} F_{1234}+\text{cyclic} \right]
\\[3mm] \label{A-F}
& = \widehat{\mathcal{A}}^{\ \rm MHV}_{6;0}\left[{R}_{146}\lr{F_{1246}+F_{1234}}+{R}_{135}F_{1235} +
{R}_{136} F_{1236}+\text{cyclic} \right]\,,
\end{align}
where $\widehat{\mathcal{A}}^{\ \rm MHV}_{6;0}$ denotes the amputated tree-level MHV super-amplitude for
$n=6$. Here we used the cyclic symmetry of the sum to make the first index of all $R$'s be 1.
The above-mentioned consistency condition leads to
\be\label{Omega-rel}
{R}_{146} =  {R}_{351} = \mathbb{P}^2 \,  {R}_{135} = \mathbb{P}^4 \,  {R}_{136}\,,
\ee
where the last relation follows from the identity \re{idfrompap}. Here $\mathbb{P}$ generates a
cyclic shift of indices $i\to i+1$ with the periodicity condition $i+6\equiv i$, or equivalently
$\mathbb{P}^6=1$. Then, it follows from \re{Omega-rel} that ${R}_{135} =\mathbb{P}^4 \,{R}_{146}$
and ${R}_{136}=\mathbb{P}^2 \,{R}_{146}$. Substituting these relations into \re{A-F} and making use
of the cyclic invariance of the sum we find
\begin{align}\label{A6-cyclic}
\widehat{\mathcal{A}}^{\ \rm NMHV}_{6;1} = \widehat{\mathcal{A}}^{\ \rm
MHV}_{6;0}\left[{R}_{146}\lr{F_{1246}+F_{1234}+F_{1345}+F_{1456}} +\text{cyclic} \right]\,,
\end{align}
The scalar box functions have infrared divergences and their expressions in dimensional
regularisation have poles in $\epsilon$ (see Appendix~A). Using their explicit expressions, we
calculate the divergent part of $\widehat{\mathcal{A}}^{\ \rm NMHV}_{6;1}$ to be
\begin{align}
\widehat{\mathcal{A}}^{\ \rm NMHV}_{6;1}\bigg|_{\epsilon\ {\rm poles}} = -\frac{ 1}{\epsilon^2}
\sum_{i=1}^n \lr{{-s_{i,i+1}}}^{-\epsilon} \times \frac12 \widehat{\mathcal{A}}^{\ \rm
MHV}_{6;0}\left[{R}_{146}  +\text{cyclic} \right]\,.
\end{align}
Comparing this relation with \re{IR-poles} we deduce the tree-level $n=6$ NMHV amplitude,
\be
\widehat{\mathcal{A}}^{\ \rm NMHV}_{6;0} = \frac12 \widehat{\mathcal{A}}^{\ \rm
MHV}_{6;0}\left[{R}_{146}  +\text{cyclic} \right] = \widehat{\mathcal{A}}^{\ \rm MHV}_{6;0}
\left[{R}_{135}  +{R}_{136}+{R}_{146} \right]\,,
\ee
where in the last relation we used the identities \re{R=R} and
\re{Omega-rel} between different $R$ coefficients. Thus, for $n=6$ we have reproduced the conjectured expression \re{gentree} for the
tree-level $n-$particle NMHV amplitude.

To finish the analysis of the $n=6$ super-amplitude, let us determine the finite part of
$\widehat{\mathcal{A}}^{\ \rm NMHV}_{6;1}$. Following \cite{dhks5}, this can be done by introducing
the ratio function
\be\label{R-finite}
\widehat{\mathcal{A}}^{\ \rm NMHV}_{n} = \widehat{\mathcal{A}}^{\ \rm MHV}_{n}\times \left[R_n^{\rm
NMHV} + O(\epsilon)\right]\,,
\ee
where $R_n^{\rm NMHV}$ is finite as $\epsilon\to 0$ and the all-loop MHV amplitude satisfies the
conjectured duality relation \re{dual-W}. We expand both sides of \re{R-finite} in powers of 't
Hooft coupling and take into account \re{A6-cyclic} and \re{dual-A} to find the ratio function for
$n=6$ as
\be\label{R6}
R_6^{\rm NMHV} = \frac12 R_{146}\left[ 1 + a V_{146}\right] + \text{cyclic}\,,
\ee
where $a=g^2 N/(8\pi^2)$ and the scalar function $V_{146}$ is given by~\cite{dhks5}
\begin{align} \notag
V_{146} & =  {F_{1246}+F_{1234}+F_{1345}+F_{1456}} - \frac12 W_{6;1}
\\ \label{V146}
& = - \ln u_1 \ln u_{2} +\frac12\sum_{k=1}^3  \big[{ \ln u_k \ln u_{k+1} + {\rm Li}_2(1-u_k)
}\big]-\frac{\pi^2}6\,.
\end{align}
where $W_{6;1}$ was defined in \re{dual-W1}. Here $u_1$, $u_2$ and $u_3$ are conformal cross-ratios
in the dual coordinates
\be
u_1=\frac{x_{13}^2x_{46}^2}{x_{14}^2x_{36}^2}\,,\qquad
u_2=\frac{x_{24}^2x_{15}^2}{x_{25}^2x_{14}^2}\,,\qquad u_3=\frac{x_{35}^2x_{26}^2}{x_{36}^2x_{25}^2}
\ee
and  the periodicity condition $u_{i+3}=u_i$ is implied. The fact that $V_{146}$ only depends on $u$
variables implies that it is invariant under conformal transformations of dual $x$ variables.

\subsubsection{Seven-point NMHV super-amplitude}

For $n=7$ the one-loop superamplitude receives contributions from three-mass, two-mass-hard,
two-mass-easy and one-mass integrals. As a consequence, the expression for
$\widehat{\mathcal{A}}_{7;1}^{\ \rm NMHV}$ can be written as
\be\label{A7-dec}
\widehat{\mathcal{A}}_{7;1}^{\ \rm NMHV} =  \widehat{\mathcal{A}}^{\ \rm MHV}_{7;0}\left[
\mathcal{I}^{\rm 3m} + \mathcal{I}^{\rm 2mh}
 + \mathcal{I}^{\rm 2me}  + \mathcal{I}^{\rm 1m}\right],
\ee
with $\mathcal{I}^{\rm 3m} = \sum \mathcal{C}^{3m} I^{3m}$ and so on.

Applying the relations \re{2mhcoeff}, \re{2mecoeff} and \re{C-1m} between the $\mathcal{C}$ coefficients and making use of the relations \re{C-I}, each contribution can be expressed as a cyclic
invariant sum of product of $R$ coefficients and scalar box $F$-functions:
\begin{itemize}
\item Three-mass contribution
\be\label{II-3m}
\mathcal{I}^{\rm 3m} = R_{146} F_{1246} + \text{cyclic}
\ee

\item Two-mass-hard contribution
\begin{align}
\mathcal{I}^{\rm 2mh} &= \lr{R_{135}+R_{251}} F_{1235} +\lr{R_{136}+R_{261}}F_{1236}+ \text{cyclic}
\end{align}

\item Two-mass-easy contribution
\begin{align}
\mathcal{I}^{\rm 2me} &= R_{461} F_{1245} +\text{cyclic} = R_{157} F_{1245}+\text{cyclic}
\end{align}
\item One-mass contribution
\begin{align}\notag
\mathcal{I}^{\rm 1m} &= \lr{R_{241}+R_{146}+R_{147}+R_{157}}F_{1234} + \text{cyclic}
\\\label{II-1m}
&=\lr{R_{241}+R_{361}+R_{351}+R_{357}}F_{1234} + \text{cyclic}
\end{align}

\end{itemize}
The two representations for $\mathcal{I}^{\rm 2me}$ and $\mathcal{I}^{\rm 1m}$ correspond to clockwise
and anti-clockwise shifts of the external legs. The consistency conditions read
\begin{align} \label{RR1}
  R_{157} &= R_{461}  = \mathbb{P}^3 \,R_{135}
\\[2mm] \notag
  R_{146}+R_{147}+R_{157} & = R_{361}+R_{351}+R_{357}= \mathbb{P}^2\,\lr{R_{146}+R_{136}+R_{135}}\,.
\end{align}
Applying $\mathbb{P}^4$ and $\mathbb{P}^5$ to the first and the second relations, respectively, and
taking into account the cyclicity condition $i+7=i$, or equivalently $\mathbb{P}^7=1$, it is easy to
see that the relations \re{RR1} are equivalent to \re{R=R}. Then, it follows from \re{R=R} and
\re{idfrompap} that
\be \label{RR2}
R_{135} = \mathbb{P}^4\, R_{157}\,,\qquad R_{136}= \mathbb{P}^2\, R_{147} \,,\qquad R_{137} =
\mathbb{P}^2\, R_{157} \,.
\ee
In addition, for $n=7$ the identity \re{genident} implies that the linear combination
\begin{align} \notag
R_{\rm tot} &= R_{135} + R_{136} + R_{137} + R_{146} + R_{147} + R_{157}
\\[3mm] \label{S}
&= R_{146} + (1+\mathbb{P}^2) R_{147} + (1+\mathbb{P}^2+\mathbb{P}^4) R_{157}
\end{align}
is cyclic invariant, $\mathbb{P}\, R_{\rm tot} = R_{\rm tot}$. Taking the sum over all cyclic shifts
of indices on both sides of the last relation we get another representation for $S$
\be\label{R-tot}
R_{\rm tot} = \frac17\left[R_{146} + 2 R_{147} + 3 R_{157} +\text{cyclic} \right]\,.
\ee

Similar to $n=6$ case, we substitute the relations \re{II-3m} -- \re{II-1m} into \re{A7-dec} and use
the cyclic invariance to make the first index of all $R$'s be 1. Then, we apply the identity \re{RR2}
to eliminate $R_{135}$, $R_{136}$ and $R_{137}$ and obtain
\begin{align}\notag
\widehat{\mathcal{A}}_{7;1}^{\ \rm NMHV} =\widehat{\mathcal{A}}^{\ \rm MHV}_{7;\,0}\big[&  R_{146}
\lr{F_{1234}+F_{1246}}+ R_{147}\lr{F_{1234}+F_{1247}+F_{1467}}
\\[3mm]
+& R_{157}\lr{F_{1234}+F_{1245}+F_{1257}+F_{1456}+F_{1567}}+\text{cyclic} \big].
\end{align}
Finally, we use the relations \re{S} to express $R_{146}$ in terms of $S$ and $R_{147}$, $R_{157}$
plus their cyclic images
\begin{align} \notag
\widehat{\mathcal{A}}_{7;1}^{\ \rm NMHV} & =\widehat{\mathcal{A}}^{\ \rm MHV}_{7;\,0}\left[ R_{\rm
tot} V_{\rm tot} +  R_{135}V_{\rm I}+   R_{147}V_{\rm II}  +\text{cyclic} \right]
\\[3mm] \label{AA-1-loop}
& = \widehat{\mathcal{A}}^{\ \rm MHV}_{7;\,0}\left[
 {\frac17  V_{\rm tot} R_{146}}+ \lr{\frac27  V_{\rm tot} +V_{\rm II}}R_{147}  + \lr{\frac37 V_{\rm tot}  +V_{\rm
I}}R_{157} +\text{cyclic} \right],
\end{align}
where in the second relation we replaced $R_{\rm tot}$ with its expression \re{R-tot}. Here the
notation was introduced for three linear combinations of scalar box functions
\begin{align} \notag
 V_{\rm tot} &=
\frac17\lr{F_{1234}+F_{1246}}+\text{cyclic}\,,
\\[1mm] \notag
V_{\rm I\phantom{I}}  &=
F_{1456}+F_{1257}+F_{1245}+F_{1567}-F_{1246}-F_{1267}-F_{2467}-F_{2457}-F_{4567}\,,
\\[3mm] \label{Vs}
V_{\rm II}&=F_{1247}+F_{1467}-F_{1246}-F_{1267}-F_{2467} \,.
\end{align}
Note that, in distinction with $V_{\rm tot}$, the functions $V_{\rm I}$ and $V_{\rm II}$ are not
cyclic invariant. We verified that the obtained one-loop NMHV super-amplitude \re{AA-1-loop}, when
expanded in powers of $(\eta_i)^4 (\eta_j)^4 (\eta_k)^4$, produces the expressions for six-gluon
one-loop NMHV amplitudes which are in agreement with the known results \cite{Bern:2004ky}.

Using expressions for scalar-box $F$ functions (see Appendix~A) it is straightforward to work out
the explicit expressions for $V_{\rm tot}$, $V_{\rm I}$ and $V_{\rm II}$ (see Eqs.~\re{VV146} and
\re{VVII} below). We find that $V_{\rm I}$ and $V_{\rm II}$ are free from infrared divergences and
are finite for $\epsilon \to 0$ while $V_{\rm tot}$ contains poles in $\epsilon$,
\begin{align}
\widehat{\mathcal{A}}_{7;1}^{\ \rm NMHV}\bigg|_{\epsilon\ \rm poles } =  -\frac{ 1}{\epsilon^2}
\sum_{i=1}^n \lr{{-s_{i,i+1}}}^{-\epsilon} \times
\widehat{\mathcal{A}}^{\ \rm MHV}_{7;\,0}R_{\rm tot}\,.
\end{align}
Comparing this relation with \re{IR-poles} we find the tree-level $n=7$ NMHV super-amplitude,
\begin{align}
\widehat{\mathcal{A}}^{\ \rm NMHV}_{7;0} &=  \widehat{\mathcal{A}}^{\ \rm MHV}_{7;0}\left[
 {\frac17    R_{146}} +  \frac27  R_{147} +  \frac37  R_{157} +\text{cyclic} \right]
 =  \widehat{\mathcal{A}}^{\ \rm MHV}_{7;0}\times \frac17\sum_{r,s,t}R_{rst}\,,
\end{align}
where we applied the identities \re{S} and \re{R-tot}. This relation is in agreement with
conjectured expression for the tree-level NMHV amplitude \re{gentree}.

According to the definition \re{R-finite}, the ratio function for $n=7$ is given to one loop by
\be\label{R7}
R_7^{\rm NMHV} = {\frac17    R_{146}}\lr{1+a V_{146}} +  \frac27  R_{147} \lr{1+a V_{\rm 147}} +
\frac37 R_{157}\lr{1+a V_{\rm 157}} +\text{cyclic}+O(a^2)\,,
\ee
where we introduced the notation for three different  combinations of scalar functions
\begin{align}\label{V147}
V_{146} = \frac12 \lr{ V_{\rm tot} - W_{7;1}}\,,\qquad V_{147} = V_{146}+\frac74 V_{\rm I}\,,\qquad
V_{157} = V_{146}+\frac76 V_{\rm II}\,,
\end{align}
with $W_{7;1}$ defined in \re{dual-W1}. The explicit expressions for these functions can be found from
\re{Vs}. We have
\be\label{VV146}
V_{146} = \frac12\left[{\rm Li_2}\lr{1-u_{1245}}-{\rm Li_2}\lr{1-u_{1246}}-\ln u_{1245}\ln u_{3467}
\right] + \text{cyclic}\,.
\ee
Notice that $V$ and $W_{7;1}$ have infrared divergences but they cancel in the difference. In a
similar manner,
\begin{align} \notag
V_{\rm I\phantom{I}} &  ={\rm Li_2}\lr{1-u_{1246}}+{\rm Li_2}\lr{1-u_{2467}} +{\rm
Li_2}\lr{1-u_{2754}}-{\rm Li_2}\lr{1-u_{1245}}
\\ \notag & \qqquad
+ \ln u_{1256} \ln u_{1745} +\ln u_{2467}\ln u_{1256} -\ln u_{2467} \ln u_{2745} -\frac{\pi^2}{6}\,,
\\ \label{VVII}
V_{\rm II} &  ={\rm Li_2}\lr{1-u_{1246}}+{\rm Li_2}\lr{1-u_{2467}} + \ln u_{1246}\ln
u_{2467}-\frac{\pi^2}{6}\,.
\end{align}
Here we use the notation for conformal ratios of dual coordinates,
\be
u_{ijkl} = \frac{x_{il}^2 x_{jk}^2}{x_{ik}^2 x_{jl}^2}\,,\qquad u_{ijkl} = u_{klij} =
\lr{u_{ijlk}}^{-1}\,.
\ee
We conclude that the functions $V_{146}$, $V_{147}$ and $V_{135}$ are finite as $\epsilon\to 0$ and,
moreover, they only depend on conformal cross-ratios of dual coordinates. We would like to stress
that this property is extremely non-trivial since infrared finiteness of a linear combination of
scalar box functions does not necessary imply that it is a function of conformal cross-ratios. We
illustrate this in Appendix~B.

Thus, $V_{146}$, $V_{147}$ and $V_{135}$ are invariant under conformal transformations of dual $x$
variables. We will argue in Sect.~5 that this property leads to dual conformality of the ratio
function $R_7^{\rm NMHV}$ given by \re{R7}.

\subsection{Infrared consistency condition}

A general expression for an arbitrary $n-$particle one-loop NMHV super-amplitude is rather involved
due to both large number of contributing cut-box diagrams and more complicated form of the
recurrence relations between two-mass-easy, one-mass and three-mass $\mathcal{C}$ coefficients,
Eqs.~\re{2mhcoeff}, \re{2mecoeff} and \re{C-1m}. To determine the tree-level NMHV amplitude from the
infrared consistency condition \re{IR-poles} it is sufficient, however, to examine the coefficient
in front of pole in $\epsilon$ with the residue depending on only one kinematical invariant, say
$\ln(-s_{12})/\epsilon$. Since not all scalar box functions contain such terms, this significantly
reduces the number of contributing terms on the right-hand side of
\re{super-amplitude-decomposition}. This  leads to the following representation
\begin{align}\label{A-new-tree}
 \widehat {\mathcal{A}}_{n;0}^{\ \rm NMHV} =\frac12 \widehat {\mathcal{A}}_{n;0}^{\ \rm MHV}\bigg[&
 2c^{\rm 1m}_{1234}+2c^{\rm 1m}_{n123}-2c^{\rm 2me}_{34n1}-c^{\rm 2mh}_{3451}-c^{\rm 2mh}_{(n-1)n13}
 +\sum_{j=5}^{n-1} c^{\rm 2mh}_{123j}- \sum_{j=6}^{n-1} c^{\rm 3m}_{34j1} -\sum_{j=5}^{n-2} c^{\rm
 3m}_{n13j}\bigg],
\end{align}
where the $c$ coefficients are related to $\mathcal{C}$ coefficients via
\be
\mathcal{C}_{r,r+1,s,t} = {c}_{r,r+1,s,t} \, \Delta_{r,r+1,s,t} \, \widehat {\mathcal{A}}_{n;0}^{\
\rm MHV}\,.
\ee
The relation \re{A-new-tree} generalizes a similar relation for tree-level gluon NMHV amplitudes found
in \cite{Bern:2004bt}.

For three-mass coefficients we find from \re{C-I} that $c^{\rm 3m}_{r,r+1,s,t} = R_{rst}$. For the
remaining coefficients we use the recurrence relations \re{2mhcoeff}, \re{2mecoeff} and \re{C-1m} to
express $c^{\rm 2me}$, $c^{\rm 2mh}$ and $c^{\rm 1m}$ as linear combinations of $R$ coefficients. In
this way, the relation \re{A-new-tree} leads to the representation for $\widehat
{\mathcal{A}}_{n;0}^{\ \rm NMHV}$ as a sum over $R$ coefficients with various indices. We recall
however that the $R$ coefficients are not independent and are related to each other by the relations
\re{idfrompap}, \re{genident} and \re{R=R}. Using these relations we can express all $R$
coefficients in terms of a basis of coefficients $R_{1st}$ with $4\le s \le t-2\le n-2$ plus their
cyclic images. We have verified by direct calculation that for $n=6,7,8,9$ the substitution of the
resulting expressions into \re{A-new-tree} yields desired result for $n$-particle tree-level NMHV
super-amplitude
\be
\widehat {\mathcal{A}}_{n;0}^{\ \rm NMHV} = \widehat {\mathcal{A}}_{n;0}^{\ \rm MHV}\times \lr{
\frac 1{n}\sum_{r,s,t=1}^n R_{rst}}\,.
\ee
It should be possible to extend this analysis for arbitrary $n$.

Finally, the ratio function \re{R-finite} takes the following form to one loop~\cite{dhks5}
\be\label{Rn}
R_n^{\rm NMHV} = \frac1{n} \sum_{s=4}^{n-2}\sum_{t=s+2}^n  m_{st} R_{1st} \lr{1 +a V_{1st}} +
\text{cyclic}\,,
\ee
where $m_{st}$ is integer combinatorial factor determined by symmetry properties of the cut-box
diagram corresponding to $R_{1st}$. Also, $V_{1st}$ is given by linear ($n-$dependent) combinations
of scalar-box functions which are infrared finite and, most importantly, dual conformal invariant.
For $n=6$ and $n=7$ the relation \re{Rn} reduces to \re{R6} and \re{R7}, respectively. Going through
the same steps as in Sect.~4.3, we have verified that \re{Rn} holds for $n=8,\, 9$.

\section{Dual superconformal symmetry}\label{DSCSy}
\label{sect-superdual}

In this section we discuss the dual superconformal symmetry of scattering amplitudes. This symmetry
was introduced in \cite{dhks5} as a generalisation of the dual conformal symmetry of MHV amplitudes.
There it was shown that the one-loop 6-point NMHV super-amplitude  exhibits dual superconformal
symmetry, in a sense which we review below. We also proposed a compact form of the $n$-particle NMHV
tree amplitude, expressed in terms of three-point dual superconformal invariants $R_{rst}$. Finally,
we formulated  the conjecture that all $\mathcal{N}=4$ SYM super-amplitudes have the property dual
superconformal symmetry.

Here we give additional evidence in favor of this hypothesis. Firstly, in subsection~\ref{Dsic} we
show that the three-mass (and the related two- and one-mass) box coefficients, obtained by means of
the generalised cut method, are indeed given by the three-point dual superconformal invariants
mentioned above. We then explain that the four-mass box coefficients from Section~\ref{gmlo} have a
similar and yet different structure, leading to a new type of four-point dual superconformal
invariants.  Further, in Section \ref{4.3} we have shown that one-loop seven-particle NMHV
super-amplitude is given by a linear combination of scalar box functions accompanied by the
coefficients $\mathcal{C}$ depending on $R_{rst}$.  In subsection~\ref{Dcii} we show that these
coefficients can be rewritten in a manifestly dual superconformal invariant way.

Recall the general form of the super-amplitude,
\be\label{recgenf}
\mathcal{A}(\Phi_1\ldots\Phi_n) =i(2\pi)^4  \delta^{(4)}(p) \delta^{(8)}(q)
\mathcal{P}_n(\lambda,\tilde\lambda,\eta)\,.
\ee
The conjecture of dual superconformal symmetry from \cite{dhks5} is formulated in terms of the
`ratio' function defined as
\be\label{faczam}
\mathcal{A}(\Phi_1\ldots\Phi_n) = \mathcal{A}^{\rm MHV}_{n}\ [R_n(\lambda,\tilde\lambda,\eta) +
O(\epsilon)]\,.
\ee
Here $\mathcal{A}^{\rm MHV}_{n}$ is the complete all-loop $n-$particle MHV super-amplitude,
including the (super) momentum conservation delta functions from \p{recgenf}. Recently, a remarkable
duality has been discovered between planar MHV amplitudes and Wilson loops in $\mathcal{N}=4$ SYM
theory \cite{am1,dks,bht}. The Wilson loop corresponding to the MHV amplitude is formulated in a
dual coordinate space. It is defined on a piecewise light-like contour $C_n$ with cusps located at
points $x_i$ related to the particle momenta via
\be\label{dualx}
p_i = x_i - x_{i+1}\,.
\ee
Then, the inherent conformal symmetry of the Wilson loop implies a surprising dual conformal
symmetry of the MHV amplitude. The conformal symmetry of the Wilson loop is broken by ultraviolet
divergences in a way controlled by an anomalous Ward identity \cite{dhks1,dhks2}. Since the
ultraviolet divergences of the Wilson loop match the infrared divergences of the scattering
amplitudes (MHV as well as non-MHV), the dual MHV amplitude has the same anomalous dual conformal
properties. This symmetry exactly predicts the form of the finite part of the (log of the) amplitude
for four and five particles, but leaves some freedom starting with six particles. The duality
conjecture goes even farther, stating that the finite parts of the Wilson loop and of the MHV
amplitude are identical for any number of points.

Let us come back to the factorised super-amplitude \p{faczam}. The conjecture made in \cite{dhks5}
claims that the anomalous dual conformal behaviour of the amplitude
$\mathcal{A}(\Phi_1\ldots\Phi_n)$ is entirely due to the divergent MHV factor $\mathcal{A}^{\rm
MHV}_{n}$, while the finite `ratio' $R_n$ is expected to be an exact dual conformal invariant. This
statement concerns the spin structures entering $R_n$, as well as all the momentum integrals
originating from the loop corrections. From the analysis in Section \ref{syinva} we know that $R_n$
is made of homogeneous polynomials in the Grassmann variables $\eta_i$ (recall \p{P-sum}). Each of
them is a combination of coefficients (spin structures) containing the $\eta$ dependence and a
function of the momenta made of loop integrals. These coefficients possess an even bigger symmetry,
they are dual superconformal invariants. We start our discussion with the latter.

\subsection{Dual superconformal invariance of the coefficients}\label{Dsic}

In order to exhibit the dual superconformal properties of the coefficients, we need to rewrite them
in dual superspace. Recall that the function $\mathcal{A}$ in \p{recgenf} is really a function of
constrained variables because it is multiplied by the (super)momentum conservation delta functions.
In other words, it is really defined only on the surface in the space of
$\lambda_i,\tilde\lambda_i,\eta_i$ described by the constraints
\be\label{conservation}
\sum_{i=1}^n  \tilde\lambda_i^{\dot\alpha}\lambda_i^{\alpha} = 0\,, \qquad \qquad \sum_{i=1}^n
\lambda_i^{\alpha} \eta_i^A = 0\,.
\ee
In \cite{dhks5} these constraints were solved by introducing a set of chiral superspace coordinates
$x_i^{ \dot\alpha\alpha},\theta_i^{A\alpha}$,
\be\label{constraints}
x_i^{\dot\alpha\alpha} - x_{i+1}^{\dot\alpha\alpha}  = \tilde\lambda_i^{\dot\alpha}
\lambda_i^{\alpha}\,, \qquad \qquad \theta_i^{A \alpha} - \theta_{i+1}^{A \alpha}  =
\lambda_i^{\alpha}\eta_i^A\,,
\ee
where we assume the cyclicity conditions $x_{n+1} \equiv x_1$ and $\theta_{n+1} \equiv \theta_1$.
These constraints imply the momentum and supercharge conservation conditions (\ref{conservation}).
The dual superconformal transformations of all variables can be deduced by assuming they act
canonically on the chiral superspace coordinates $x_i,\theta_i$ and are compatible with the
constraints (\ref{constraints}). The details can be found in \cite{dhks5}, here we just recall a few
basic points.

The dual conformal properties of all objects formulated in dual superspace are most easily verified
by performing conformal inversion. It acts on the dual superspace coordinates as follows:
\begin{align}\notag
I[x_{\alpha\dot\beta}] & = \frac{x_{\beta\dot\alpha}}{x^2} \equiv (x^{-1})_{\beta\dot\alpha}\,, &&
\hspace*{-20mm} I\left[\theta_{i}^{A\, \alpha}\right] = (x_i^{-1})^{\dot\alpha\beta}\theta_{i\,
\beta}^A\,,
\\[2mm]\label{conin}
I\left[\lambda_i^\alpha\right] & = (x_i^{-1})^{\dot\alpha\beta}\lambda_{i\,\beta}\,,
  && \hspace*{-20mm} I[\tilde\lambda_{i\,\dot\alpha}] = (x_{i+1}^{-1})_{\alpha\dot\beta}
  \tilde\lambda_{i+1}^{\dot\beta}\,.
\end{align}
The transformations of $x$ and $\theta$ are standard, while those of $\lambda$ and $\tilde\lambda$
are derived from the bosonic constraint in \p{constraints}. \footnote{The dual conformal
transformations of the `super-momenta' $\eta_i$ are inhomogeneous \cite{dhks5}, but they are not
necessary for our discussion here.} With the help of these rules it is very easy to see that various
Lorentz invariant contractions of spinors $\lambda$ (or $\tilde\lambda$) and dual space vectors
$x_{ij}$ are conformally covariant, for example
\begin{equation}\label{exconin}
    I[x^2_{ij}] = \frac{x^2_{ij}}{x^2_i x^2_j}\,, \qquad I \Big[\vev{i\ i+1}\Big] =  (x^2_{i})^{-1}\ {\vev{i\ i+1}}\,, \qquad I\Big[ \vev{i|x_{ij}x_{jk}|k} \Big] =  \frac{ \vev{i|x_{ij}x_{jk}|k}}{x^2_i x^2_j x^2_k}\,.
\end{equation}

Now, let us examine the  dual superconformal properties of the main building block \p{Omega} of the
three-, two- and one-mass coefficients from Section \ref{section-unitarity-NMHV},
\be\label{Omega'}
R_{tsu} =  -\frac{\vev{s-1\,s}\vev{u-1\,u}\
\delta^{(4)}\bigl(\Xi_{tsu}\bigr)}{x_{su}^2\vev{t|x_{ts}x_{su}|u-1}\vev{t|x_{ts}x_{su}|u}
\vev{t|x_{tu}x_{us}|s-1}\vev{t|x_{tu}x_{us}|s}}\,.
\ee
The various bosonic factors here are of the types shown in \p{exconin}, so they are covariant under
conformal inversion. The Grassmann dependence resides in the linear combination of $\eta$'s
\p{iniforxi}, which can be rewritten in terms of the dual superspace coordinates \p{constraints} as
follows:
\begin{eqnarray}
  \Xi_{tsu} &=& \sum_{u}^{t-1} \eta_i \vev{i|x_{us}x_{st} |t} +
 \sum_{r}^{s-1}\eta_i \vev{i|x_{su}x_{ut} |t} \nonumber\\
  &=& x^2_{su} \vev{t |\theta_t} + \vev{t| x_{ts} x_{su} |\theta_u} + \vev{t| x_{tu} x_{us}| \theta_s}\,. \label{rewrxi}
\end{eqnarray}
Applying once again the rules \p{conin}, we can easily show that this combination is dual conformal
covariant,
\begin{equation}\label{XiI}
    I[\Xi_{tsu}] = \frac{\Xi_{tsu}}{x^2_t x^2_s x^2_u}\,.
\end{equation}
Then, combining this with the transformations of the bosonic factors in \p{rewrxi}, we see that
$R_{tsu}$  \p{Omega'} is indeed a dual conformal invariant.

In fact, it is invariant not only under dual conformal, but also superconformal transformations. One
of the generators of dual Poincar\'e supersymmetry, $Q_{\alpha A}$, acts as a shift of the chiral
dual superspace coordinates, $\delta\theta_i^{A \alpha} = \epsilon_i^{A \alpha}$, while leaving the
Grassmann variables $\eta_i$ invariant. \footnote{The dual Poincar\'e supersymmetry algebra with
generators $Q, \bar Q$ should not be confused with the original supersymmetry \p{q} of the amplitude
with generators $q, \bar q$.} The invariance of $\Xi_{rst}$ is obvious in its initial form
\p{iniforxi}, and is easy to show in the form \p{rewrxi} due to the identity $x^2_{su} + x_{ts}
x_{su} + x_{tu} x_{us} = 0$. The other generators of Poincar\'e supersymmetry, $\bar
Q^A_{\dot\alpha}$, acts on the bosonic coordinates $x$ and $\tilde\lambda$. To show the invariance
of \p{Omega'} is not that simple, so we refer the reader to \cite{dhks5} for the explanations.
Combining these two generators with conformal inversion, we can obtain the rest of the
$\mathcal{N}=4$   superconformal algebra.

\begin{figure}
\psfrag{g}[cc][cc]{$r$}\psfrag{h}[cc][cc]{$s-1~~$}\psfrag{a}[cc][cc]{$s$}\psfrag{b}[cc][cc]{$t-1$}
\psfrag{c}[cc][cc]{$t$}\psfrag{d}[cc][cc]{$u-1$}\psfrag{e}[cc][cc]{$u$}\psfrag{f}[cc][cc]{$r-1$}
%
\centerline{{\epsfysize5.5cm \epsfbox{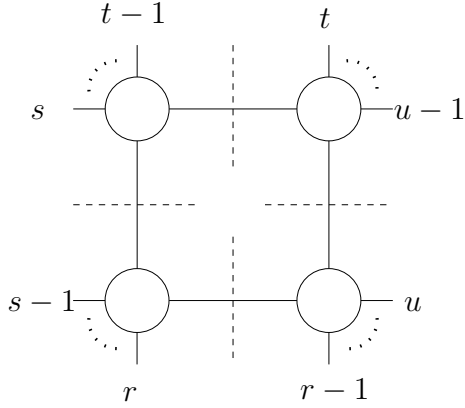}}} \caption[]{\small The configuration contributing
to the four-mass coefficient $\mathcal{C}^{\rm 4m}_{r,s,t,u}$. The empty vertices are MHV
super-amplitudes. } \label{Fig:4mass}
\end{figure}

The four-mass coefficients were obtained in Section \ref{gmlo} as the most straightforward
application of the quadruple cut technique, see \p{P-4m}. According to \p{P-4m-gen} and the
discussion afterwards, they cannot contribute to super-amplitudes of the MHV or NMHV type. Let us
examine the simplest of them, the first term in \p{P-4m-gen}, which contributes to NNMHV amplitudes
(see Fig.~\ref{Fig:4mass}). It is obtained by substituting all the vertex factors
$\mathcal{P}_{n_i+2;0}$ in \p{P-4m} with the bosonic factors of the  MHV tree amplitudes, see
\p{seta0}. The Grassmann integration has already been done in \p{P-4m}, so we just need to collect
all bosonic factors and simplify them. The calculation is very similar to that of the three-mass
coefficients from Section \ref{3mcose}, and we obtain
\begin{equation}\label{weobtain}
     \mathcal{P}_{n;1}^{\rm 4m} = \frac{1}{\prod_{i=1}^n \vev{i\, i+1}}\ \frac{x^2_{rt}x^2_{su}}{2 x^2_{rs}x^2_{tu}}[x^2_{rs}x^2_{tu} + x^2_{ru}x^2_{st} - x^2_{rt}x^2_{su}] \frac12
\sum_{\mathcal{S}_\pm} \hat R_{l_3;tsu}\ \hat R_{l_4;urt}\,.
\end{equation}
Here we have introduced the new superconformal invariant
\begin{equation}\label{newome}
    \hat R_{l_3;tsu} = -\frac{\vev{s-1\,s}\vev{u-1\,u}\  \delta^{(4)}\bigl(\hat\Xi_{l_3;tsu} \bigr)}{x_{su}^2\vev{l_3|x_{tu}x_{us}|s-1}\vev{l_3|x_{tu}x_{us}|s}
\vev{l_3|x_{ts}x_{su}|u-1}\vev{l_3|x_{ts}x_{su}|u}}
\end{equation}
with
\begin{equation}\label{hatxi}
    \hat\Xi_{l_3;tsu} = x^2_{su} \vev{l_3 |\theta_r} + \vev{l_3| x_{ts} x_{su} |\theta_u} + \vev{l_3| x_{tu} x_{us}| \theta_s} \,,
\end{equation}
and similarly for $\hat R_{l_4;urt}$. It looks very similar to the invariant $R_{tsu}$ \p{Omega'},
being made of the same triplet of dual superspace points, $(x_{t,s,u},\, \theta_{t,s,u})$. The only
difference is that in \p{newome} and in \p{hatxi} we use the `internal' spinor variable
$\lambda_{l_3}$ to obtain Lorentz invariant projections, while in \p{Omega'} and \p{rewrxi} this is
done with the `external' spinor variable $\lambda_{t}$, associated with the momentum $p_t = x_t -
x_{t+1}$. In principle, the  internal spinors $\lambda_{l_i}$ ($i=1,2,3,4$) are determined from the
loop momenta $(l_i)^{\dot\alpha\alpha} = \tilde\lambda_{l_i}^{\dot\alpha}\lambda_{l_i}^{\alpha}$, up
to an arbitrary scale. Since the expression \p{newome} is homogeneous in $\lambda_{l_3}$, this scale
drops out. Further, the loop momenta $l_i$ themselves can be solved for in terms of the external
momenta $K_i$ from the kinematical constraints \p{cutcond}. Thus, we may say that $\hat R_{l_3;tsu}$
is effectively a function of the external momenta only, but making this statement explicit is a
non-trivial task.

Nevertheless, we are in a position to claim that $\hat R_{l_3;tsu}$ is a dual conformal invariant.
First, we need to find out how the internal spinors $\lambda_{l_i}$ transform under inversion.
Remembering that $K_1 = x_{rs}$, $K_2 = x_{st}$, $K_3 = x_{tu}$ and $K_4 = x_{ur}$,  we can solve
the momentum conservation constraints \p{cutcond} by introducing an extra point in dual space,
$x_0$, and writing
\begin{equation}\label{extrax}
    l_1 = x_{0r}\,, \quad l_2 = x_{0s}\,, \quad l_3 = x_{0t}\,, \quad l_4 = x_{0u}\,.
\end{equation}
The momenta $l_i$ still satisfy the on-shell conditions  $l^2_i=0$, which are solved through the
internal spinor variables $(l_i)^{\dot\alpha\alpha} =
\tilde\lambda_{l_i}^{\dot\alpha}\lambda_{l_i}^{\alpha}$. These relations establish a link between
$\lambda_{l_i}$ and the points in dual space. For example, consider the relation
$(l_3)^{\dot\alpha\alpha} = \tilde\lambda_{l_3}^{\dot\alpha}\lambda_{l_3}^{\alpha} =
(x_{0t})^{\dot\alpha\alpha}$. Performing a conformal inversion on it, we can derive the
transformation of  $\lambda_{l_3}$, analogous to that of $\lambda_i$ in \p{conin}, \footnote{Due to
the constraint $\lambda_{l_3}^\alpha (x_{0t})_{\alpha\dot\alpha} = 0$ one can replace the matrix
$(x_t^{-1})^{\dot\alpha\beta} = (x_t)^{\dot\alpha\beta}/x^2_t$ in \p{tral3} by
$(x_0)^{\dot\alpha\beta}/x^2_t$.}
\begin{equation}\label{tral3}
    I\left[\lambda_{l_3}^\alpha\right] = (x_t^{-1})^{\dot\alpha\beta}\lambda_{{l_3}\,\beta}\,.
\end{equation}
We see that the internal spinor $\lambda_{l_3}$ transforms in exactly the same way as the external
$\lambda_{t}$, therefore the dual conformal invariance of $\hat R_{l_3;tsu}$ \p{newome} can be
proven in the same way as for $R_{tsu}$ \p{Omega'}.

We can say that the three-mass (and the related two- and one-mass) coefficients are given in terms
of the simpler, manifestly three-point invariants $R_{tsu}$ because they contain at least one
three-particle vertex. This leads to kinematical constraints which relate the internal spinor
variables at this vertex to the external. Precisely this  external spinor is used in  $R_{tsu}$ to
project at the first point. In the non-degenerate four-mass case the internal spinors decouple from
the external, which explains the more complicated structure of $\hat R_{l_3;tsu}$.

Coming back to the four-mass coefficient \p{weobtain}, besides the two dual conformal invariants
$\hat R$, we see a prefactor made of dual space `distances' $x^2_{ij}$. It is a dual conformal
covariant with the necessary conformal weight which turns the accompanying (finite) four-mass box
integral into a dual conformal invariant.

\subsection{Dual conformal invariance of the integrals}\label{Dcii}

Let us now examine the conformal properties of the scalar box integrals entering
\re{super-amplitude-decomposition}. In the dual variables \re{constraints}, they are functions of
the $x$'s only. If one did not pay attention to the convergence properties of the scalar box
integrals, they would be formally dual conformal covariant in four dimension. However, these
functions suffer from infrared divergences and, in distinction with the coefficients $\mathcal{C}$,
they are well only defined in $D=4-2\epsilon$ dimensions with $\epsilon\neq 0$. This implies that
their dual conformal symmetry is broken by infrared singularities.

We notice that the infrared divergences of the scattering amplitudes have a universal form
independent of the helicity configuration of the external particles. As a result, writing down the
NMHV super-amplitude in the factorized form \re{R-finite}, we absorb all infrared poles in
$\epsilon$ into the MHV super-amplitude and define the infrared finite ratio function $R_n^{\rm
NMHV}$. To one-loop order, this function admits an expansion similar to
\re{super-amplitude-decomposition}. Taking into account the relations between the $\mathcal{C}$
coefficients, the ratio function is given by a sum \re{Rn} over independent (dual superconformal
invariant) $R_{rst}$ coefficients multiplied by linear combinations $V_{rst}$ of scalar box
functions. A characteristic feature of these combinations is that they are finite as $\epsilon\to
0$, while each term separately has poles in $\epsilon$. The following two questions arise:
\begin{itemize}
\item Does the condition of infrared finiteness fix the form of the linear combinations of
scalar box integrals uniquely (up to cyclic shift of indices)?

\item Is dual conformal symmetry restored in the infrared finite combinations of scalar box
integrals?
\end{itemize}
It turns out that for $n=6$ both questions have an affirmative answer leading to $V_{146}$,
Eq.~\re{V146}. The situation changes however for $n\ge 7$. We find that for $n=7$ there exist three
linear combinations of three-mass, two-mass-hard, two-mass-easy and one-mass scalar boxes that are
finite as $\epsilon\to 0$. They are given by $V_{\rm I}$, $V_{\rm II}$ and $F$ defined in \re{Vs}
and \re{F}, respectively. It is straightforward to verify (see Eqs.~\re{VVII} and \re{KF}) that only
the first two functions are dual conformal invariant. We also verified that both questions have
negative answer for $n=8$.\footnote{In this case, we also encounter four-mass integrals which are
infrared finite and dual conformal simultaneously. These integrals contribute starting with the
NNMHV super-amplitude.}

The fact that the ratio function $R_n^{\rm NMHV}$ is finite as $\epsilon\to 0$ implies that its
expansion may involve all infrared finite combinations of scalar box integrals. Our analysis for
$n=6,7,8$ shows that $R_n^{\rm NMHV}$  receives contribution from the dual conformal combinations
only. This result is in agreement with the conjecture of \cite{dhks5} that the ratio function
$R_n^{\rm NMHV}$ should be equal, at one-loop, to a linear combination of $R_{rst}$ coefficients
accompanied by conformal invariant functions $V_{rst}$ of dual $x$ coordinates.

\section*{Acknowledgements}

We would like to thank Babis Anastasiou, Zvi Bern, Ruth Britto, Lance Dixon, Paul Heslop, David
Kosower, Zoltan Kunszt and Radu Roiban for stimulating discussions. G.K and E.S. are grateful to the
Center for Theoretical Studies, ETH Z\"urich for hospitality during the final stage of this work.
This research was supported in part by the French Agence Nationale de la Recherche under grant
ANR-06-BLAN-0142.

\section*{Appendices}

\appendix

\setcounter{section}{0} \setcounter{equation}{0}
\renewcommand{\theequation}{\Alph{section}.\arabic{equation}}

\section{Scalar box integrals}
\label{sect-integrals}

Following \cite{bddk94,Bern:2004ky}, we define scalar box $F$ functions as
\begin{align}\label{box}
-i (4\pi)^{2-\epsilon}  \int  \frac{d^{4-2\epsilon}l}{(2\pi)^{4-2\epsilon}} \frac1{l^2 (l+K_1)^2
(l+K_1+K_2)^2(l-K_4)^2 } = r_\Gamma \frac{F(K_1,K_2,K_3,K_4)}{\Delta(K_1,K_2,K_3,K_4)}\,,
\end{align}
where $\Delta(K_1,K_2,K_3,K_4)$ is given by
\begin{align}
 \Delta(K_1,K_2,K_3,K_4) = -  {2\sqrt{\det
\|S\|}} \,,
\end{align}
the symmetric $4\times 4$ matrix $S$ has components
\begin{align}
S_{ij} = -\ft12 (K_i+\ldots + K_{j-1})^2\,, \quad (i\neq j)\,,\qquad S_{ii}=0\,,
\end{align}
with $i,j$ defined  modulo 4, and the normalization factor is
\begin{align}\label{c-G}
r_\Gamma = {c}_\Gamma(4\pi)^{2-\epsilon}\,,\qquad
{c}_\Gamma=\frac1{(4\pi)^{2-\epsilon}}\frac{\Gamma(1+\epsilon)\Gamma^2(1-\epsilon)}{\Gamma(1-2\epsilon)}\,.
\end{align}
The 4-vectors $K_{1,2,3,4}$ are sums of external on-shell momenta $p_i$ of the $n-$point amplitude.
They can be specified by four ordered indices $r < s < t <u $ (mod $n$) as follows
\be\label{K-dual}
K_1 = \sum_{r}^{s-1} p_i =x_{sr}\,,\quad K_2 = \sum_s^{t-1} p_i=x_{ts}\,,\quad K_3 = \sum_{t}^{u-1}
p_i=x_{ut}\,,\quad K_4=\sum_u^{s-1} p_i=x_{su}\,,
\ee
where we introduced the dual variables $p_i=x_i-x_{i+1}$ and $x_{ij}\equiv x_i -x_j$. This suggests
to use the shorthand notations
\begin{align}\label{FF}
F_{rstu} \equiv F(K_1,K_2,K_3,K_4)\,,\qquad \Delta_{rstu} \equiv \Delta(K_1,K_2,K_3,K_4)\,,
\end{align}
with $F_{rstu}$ and $\Delta_{rstu}$ being symmetric in any pair of indices. The expression for
$F_{rstu}$ takes a different form depending on whether $K_i^2=0$ or $K_i^2\neq 0$. In this way, we
obtain (all indices are defined modulo $n$):

\noindent for one-mass function
\begin{align}\notag
F_{i-3,\,i-2,\,i-1,\,i}  &= -\frac1{\epsilon^2} \bigg[ (-x_{i-3,i-1})^{-\epsilon} +
(-x_{i-2,i})^{-\epsilon} - (-x_{i,i-3})^{-\epsilon} \bigg]
\\ \notag
&+ {\rm Li}_2 \lr{1-\frac{x_{i,i-3}^2}{x_{i-3,i-1}^2}}+ {\rm
Li}_2\lr{1-\frac{x_{i,i-3}^2}{x_{i-2,i}^2}}+\frac12\ln^2\lr{\frac{x_{i-3,i-1}^2}{x_{i-2,i}^2}}+\frac{\pi^2}{6}
\\
&\equiv F_{n;i}^{\rm 1m} \label{box-1m}
\end{align}
for easy two-mass
\begin{align} \notag
F_{i-1,\,i,\,i+r,\,i+r+1} &= -\frac1{\epsilon^2}\bigg[
(-x_{i-1,i+r})^{-\epsilon}+(-x_{i,i+r+1})^{-\epsilon}-(-x_{i,i+r})^{-\epsilon}-(-x_{i+r+1,i-1})^{-\epsilon}\bigg]
\\ \notag
&+{\rm Li}_2\lr{1-\frac{x_{i,i+r}^2}{x_{i-1,i+r}^2}}+{\rm
Li}_2\lr{1-\frac{x_{i,i+r}^2}{x_{i,i+r+1}^2}}+{\rm Li}_2\lr{1-\frac{x_{i+r+1,i-1}^2}{x_{i-1,i+r}^2}}
\\ \notag
& +{\rm Li}_2\lr{1-\frac{x_{i+r+1,i-1}^2}{x_{i,i+r+1}^2}}-{\rm Li}_2\lr{1-\frac{x_{i,i+r}^2
x_{i+r+1,i-1}^2}{x_{i-1,i+r}^2 x_{i,i+r+1}^2}} +
\frac12\ln^2\lr{\frac{x_{i-1,i+r}^2}{x_{i,i+r+1}^2}}
\\
&\equiv F_{n;r;i}^{\rm 2m\ e} \label{box-2me}
\end{align}
for hard two-mass
\begin{align} \notag
F_{i-2,\,i-1,\,i,\,i+r} &= -\frac1{\epsilon^2}\bigg[
(-x_{i-2,i})^{-\epsilon}+(-x_{i-1,i+r})^{-\epsilon}-(-x_{i,i+r})^{-\epsilon}-(-x_{i+r,i-2})^{-\epsilon}
\bigg]
\\ \notag
&-\frac1{2\epsilon^2}\frac{(-x_{i,i+r})^{-\epsilon}(-x_{i+r,i-2})^{-\epsilon}}{(-x_{i-2,i})^{-\epsilon}}
+ \frac12\ln^2\lr{\frac{x_{i-2,i}^2}{x_{i-1,i+r}^2}}
\\ \notag
& +{\rm Li}_2\lr{1-\frac{x_{i,i+r}^2}{x_{i-1,i+r}^2}}+{\rm
Li}_2\lr{1-\frac{x_{i+r,i-2}^2}{x_{i-1,i+r}^2}}
\\
&\equiv F_{n;r;i}^{\rm 2m\ h}  \label{box-2mh}
\end{align}
for three mass
\begin{align} \notag
F_{i-1,\,i,\,i+r,\,i+r+r'} = \hspace*{-20mm} &
\\ \notag &
 -\frac1{\epsilon^2}\bigg[ (-x_{i-1,i+r}^2)^{-\epsilon}
+(-x_{i,i+r+r'}^2)^{-\epsilon}-(-x_{i,i+r}^2)^{-\epsilon}-(-x_{i+r,i+r+r'}^2)^{-\epsilon}
-(-x_{i+r+r',i-1}^2)^{-\epsilon}\bigg]
\\& \notag
-\frac1{2\epsilon^2}\frac{(-x_{i,i+r}^2)^{-\epsilon}(-x_{i+r,i+r+r'}^2)^{-\epsilon}}{(-x_{i,i+r+r'}^2)^{-\epsilon}}
-\frac1{2\epsilon^2}\frac{(-x_{i+r,i+r+r'}^2)^{-\epsilon}(-x_{i+r+r',i-1}^2)^{-\epsilon}}{(-x_{i-1,i+r}^2)^{-\epsilon}}
+ \frac12\ln^2\lr{\frac{x_{i-1,i+r}^2}{x_{i,i+r+r'}^2}}
\\ \notag
& +{\rm Li}_2\lr{1-\frac{x_{i,i+r}^2}{x_{i-1,i+r}^2}}+{\rm
Li}_2\lr{1-\frac{x_{i+r+r',i-1}^2}{x_{i,i+r+r'}^2}}-{\rm Li}_2\lr{1-\frac{x_{i,i+r}^2
x_{i+r+r',i-1}^2}{x_{i-1,i+r}^2 x_{i,i+r+r'}^2}}
\\
&\equiv F^{\rm 3m}_{n;r,r';i}\,.  \label{box-3m}
\end{align}
The corresponding expressions for $\Delta_{r,r+1,st}$ are
\be
\Delta_{r,r+1,s,t} = -\frac12 \big[x_{rs}^2 x_{r+1\, t}^2- x_{rt}^2 x_{r+1\, s}^2 \big]\,.
\ee


\section{Infrared finiteness versus dual conformality}

The super-amplitude \re{super-amplitude-decomposition} is given by the sum over scalar box integrals
\re{I} accompanied with $\mathcal{C}$ coefficients. Introducing the dual coordinates for external
momenta \re{K-dual} and for the loop momentum, $l=x_r-x_0$, one finds that the scalar box integral
\re{I} takes the form $\sim  \int {d^{4-2\epsilon} x_0} \lr{x_{r0}^2 x_{s0}^2 x_{t0}^2
x_{u0}^2}^{-1}$. If this integral was well-defined in four dimensions (for $\epsilon=0$), it would
be covariant under conformal transformations of $x$ variables and, as a consequence, the functions
$F_{rstu}$, Eq.~\re{FF}, would be dual conformal invariant.  This is indeed the case for the
four-mass scalar function. The remaining (three-mass, two-mass-hard, two-mass-easy and one-mass)
scalar functions have infrared divergences and require regularization. In the dimensional
regularization with $D=4-2\epsilon$ their conformal symmetry is broken for $\epsilon\neq 0$.

There exist linear combinations of divergent scalar box functions which remain finite as $\epsilon
\to 0$.  Since infrared divergences cancel in the sum of scalar box functions one may expect that
the dual conformal invariance gets restored in such combinations. We demonstrated in Sect.~4.3, this
is indeed the case at $n=6$ for $V_{146}$, Eq.~\re{V146}, and at $n=7$ for $V_{135}$, $V_{146}$ and
$V_{147}$, Eqs.~\re{V147}. These linear combinations are exceptional since, as we will show in a
moment, infrared finiteness does not automatically implies  dual conformality.

Let us examine the following linear combination of one-mass and two-mass-hard scalar functions
defined for $n=7$
\begin{equation}\label{F}
F = -F_{4567} + F_{3567} - F_{3457} + F_{3456}\,.
\end{equation}
It depends on the dual coordinates $x_3,x_4,x_5,x_6,x_7$. Using expressions for the scalar
functions, Eqs.~\re{box-1m} and \re{box-2mh}, we verify that each term in the right-hand side of
\re{F} contains infrared divergences but poles in $\epsilon$ cancel in their sum.  The resulting
expression for $F$ is finite as $\epsilon \to 0$:
\begin{align}
F =&\frac{1}{2}\Bigl(\ln^2 \frac{x_{35}^2}{x_{46}^2} - \ln^2 \frac{x_{35}^2}{x_{47}^2} - \ln
\frac{x_{35}^2}{x_{57}^2} \ln \frac{x_{37}^2}{x_{57}^2} - \ln^2 \frac{x_{46}^2}{x_{57}^2} + \ln
\frac{x_{37}^2}{x_{35}^2} \ln \frac{x_{57}^2}{x_{35}^2} + \ln^2
\frac{x_{57}^2}{x_{36}^2}\Bigr) \notag \\
&+{\rm Li}_2\Bigl(1-\frac{x_{35}^2}{x_{36}^2}\Bigr) +{\rm
  Li}_2\Bigl(1-\frac{x_{36}^2}{x_{35}^2}\Bigr) +{\rm
  Li}_2\Bigl(1-\frac{x_{37}^2}{x_{36}^2}\Bigr) +{\rm
  Li}_2\Bigl(1-\frac{x_{36}^2}{x_{46}^2}\Bigr) \notag \\
&-{\rm
  Li}_2\Bigl(1-\frac{x_{37}^2}{x_{47}^2}\Bigr) -{\rm
  Li}_2\Bigl(1-\frac{x_{47}^2}{x_{46}^2}\Bigr) -{\rm
  Li}_2\Bigl(1-\frac{x_{47}^2}{x_{57}^2}\Bigr) -{\rm
  Li}_2\Bigl(1-\frac{x_{57}^2}{x_{47}^2}\Bigr).
\end{align}
To verify its dual conformality we apply the conformal boost $K^\mu = \sum_{i=1}^6 (2x_i^\mu
(x_i\partial_{x_i})-x_i^2\partial_{x_i}^\mu)$  to both sides of this relation
\begin{align}\label{KF}
K^\mu F =x_3^\mu \Bigl( \frac{x_{47}^2}{x_{37}^2-x_{47}^2}\ln \frac{x_{37}^2}{x_{47}^2} -
\frac{x_{46}^2}{x_{36}^2-x_{46}^2}\ln \frac{x_{46}^2}{x_{36}^2}\Bigr) +x_4^\mu
\Big(\frac{x_{36}^2}{x_{36}^2-x_{46}^2} \ln \frac{x_{36}^2}{x_{46}^2} -
\frac{x_{37}^2}{x_{37}^2-x_{47}^2} \ln
\frac{x_{37}^2}{x_{47}^2}\Bigr) \notag \\
+x_6^\mu \Big(\frac{x_{37}^2}{x_{36}^2-x_{37}^2} \ln \frac{x_{36}^2}{x_{37}^2} -
\frac{x_{47}^2}{x_{46}^2-x_{47}^2} \ln \frac{x_{46}^2}{x_{47}^2}\Bigr) +x_7^\mu
\Bigl(\frac{x_{46}^2}{x_{46}^2-x_{47}^2} \ln \frac{x_{46}^2}{x_{47}^2}
-\frac{x_{36}^2}{x_{36}^2-x_{37}^2} \ln \frac{x_{36}^2}{x_{37}^2} \Bigr),
\end{align}
so $K^\mu F \neq 0$ and, as a consequence, $F$ is not conformal invariant. Six other infrared
finite, non-conformal combinations can be obtained by rotating the seven points $\{x_1,...,x_7\}$
cyclically.


\end{document}